\documentclass[12pt]{iopart}

\usepackage{float}
\usepackage{graphicx}
\usepackage{hyperref}
\usepackage{iopams}
\usepackage{subcaption}
\usepackage{xcolor}

\begin{document}

\title{Non-positive energy quasidistributions in coherent collision models}

\author{Marco Pezzutto$^1$\footnote{Author to whom any correspondence should be addressed.}, Gabriele De Chiara$^{2,3}$ and Stefano Gherardini$^{1,4}$}

\address{$^1$ Istituto Nazionale di Ottica del Consiglio Nazionale delle Ricerche (CNR-INO), Largo Enrico Fermi 6, I-50125 Firenze, Italy}
\address{$^2$ Centre for Quantum Materials and Technology, School of Mathematics and Physics, Queen’s University Belfast, Belfast BT7 1NN, United Kingdom.}
\address{$^3$Física Teòrica: Informació i Fenòmens Quàntics, Departament de Física, Universitat Autònoma de Barcelona, 08193 Bellaterra, Spain.}
\address{$^4$European Laboratory for Non-linear Spectroscopy, Università di Firenze, I-50019 Sesto Fiorentino, Italy}
\eads{\mailto{marco.pezzutto@ino.cnr.it}, \mailto{gabriele.dechiara@uab.cat} and \mailto{stefano.gherardini@ino.cnr.it}}

\begin{abstract}
We determine the Kirkwood-Dirac quasiprobability (KDQ) distribution associated to the stochastic instances of internal energy variations for the quantum system and environment particles in coherent Markovian collision models. In the case the interactions between the quantum system and the particles do not conserve energy, the KDQ of the non-energy-preserving stochastic work is also derived. These KDQ distributions can account for non-commutativity, and return the unperturbed average values and variances for a generic interaction-time, and generic local initial states of the quantum system and environment particles. Using this nonequilibrium-physics approach, we certify the conditions under which the collision process of the model exhibits quantum traits, and we quantify the rate of energy exchanged by the quantum system by looking at the variance of the KDQ energy distributions. Finally, we propose an experimental test of our results on a superconducting quantum circuit implementing a qubit system, with microwave photons representing the environment particles.    
\end{abstract}

\vspace{2pc}

\date{\today}

\section{Introduction}
\label{sec:introduction}

The modelling and characterisation of open quantum systems is of paramount importance for the description of complex quantum systems in contact with their environment, the assessment and protection of noisy quantum devices, and the study of the emergence of the fundamental laws of thermodynamics in out-of-equilibrium quantum systems. 
In collision models (see, for instance, \cite{CampbellEPL2021,CusumanoReviewEntropy,CiccarelloReview,Ciccarello2017}) the environment is made of simple identical units, for example qubits or quantum harmonic oscillators, which interact for a short time with the system and are then discarded. These models play a fundamental and increasingly important role thanks to their versatility, theoretical simplicity, and their ability to describe an enormous class of dissipative evolutions, including all Markovian ones~\cite{GiovannettiPRL2012,CattaneoPRL2021}. Besides, they have proven to be valuable tools for approaching  scenarios in open quantum systems which are usually challenging to tackle with typical techniques, such as non-Markovian dynamics~\cite{CiccarelloGiovannetti,Ciccarello2013}, and when system-environment correlations play a significant role~\cite{McCloskeyPhysRevA2014}.

Collision models have been the focus of several investigations in non-equilibrium dynamics and thermodynamics~\cite{DeChiaraNJP2018,CusumanoPRA2018,SeahPRE2019,PiccionePRA2021,ShaghaghiPRE2022,CarolloQST2024,PezzuttoNJP2016} including cases in which auxiliary systems are initially endowed with some form of quantum coherence or correlation~\cite{RodriguesPRL2019,DeChiaraPRR2020,HammamNJP2022,HammamPRR2024,CusumanoNJP2024}, showing a quantum advantage in battery charging by repeated interactions~\cite{MorroneQST2023,SalviaPRR2023} and in thermometry~\cite{SeahPRL2019,ShuPRA2020,AlvesPRA2022}. Collision models have also been employed to study heat and work distributions in a driven open quantum system developing non-Markovian dynamics~\cite{Zou2024}, using the non-demolition quasiprobability approach~\cite{Paolo2016,Paolo2021,Paolo2022}.

In this paper, we make use of Kirkwood-Dirac quasiprobabilities (KDQs)~\cite{kirkwood1933quantum,dirac1945analogy,yunger2018quasiprobability,DeBievrePRL2021,LostaglioQuantum2023,SantiniPRB2023,BudiyonoPRAquantifying,FrancicaPRE2023,wagner2023quantum,GherardiniTutorial,ArvidssonShukur2024review,SagarQSL2024} to perform a complete analysis of the non-equilibrium thermodynamics of a collision model. KDQs allow us to assess the impact of initial quantum coherence on the thermodynamics in a multi-time process, which is usually not possible with other approaches such as the two-point measurement scheme, which causes initial coherences to be erased because of the measurement process~\cite{GherardiniTutorial}. While KDQs have been already applied to the study of non-equilibrium thermodynamics of quantum systems, our approach based on collision models is able to characterize non-equilibrium features of a collision model in the case the states of both the system and each environment particle contain quantum coherence. In addition, as discussed below, KDQs also proved useful when determining not just the average of the distributions associated to relevant thermodynamic quantities, but also higher moments, in particular the second moment and the variance.

In the following we provide the distributions of the stochastic instances of internal energy variations of system and environmental auxiliary system, and of non-energy-preserving work, and prove the validity of the $1^{\rm st}$ law of thermodynamics. In case the energy-preserving condition is valid, we can also determine the distributions of the stochastic instances of the so-called coherent work and incoherent heat, according to the definition in~\cite{RodriguesPRL2019}.

We show that the corresponding KDQ distributions can account for non-commutativity, and show the statistical moments of the distributions (including average and variances) for a generic interaction-time and for generic local initial states of the system and the environmental auxiliary system. Our analysis, stemming from the dictates of stochastic thermodynamics, offers the possibility of  going naturally beyond the small interaction-time limit, and at the same time giving results beyond the averages values.

We certify the conditions under which the collision process of the model exhibits quantum traits. To attain this task, as a quantumness criterion, we use the negativity of the real part of a KDQ, as well as the fact that the imaginary part of a KDQ is different from zero. Moreover, we quantify the rate of energy exchange between the system and the environmental auxiliary systems, using the variances of the KDQ energy distributions.

Our general results are showcased on a qubit-qubit case study whereby both the quantum system and the environmental auxiliary system are modelled by a qubit; in such a case, analytical formulas can be provided for almost all quantities of interest. Finally, we suggest experimental tests of our results on concrete physical systems, in particular on microwave photonics with superconducting quantum circuits, that allow for the implementation of a Jaynes-Cummings  interaction between a superconducting qubit and microwave photons.

\section{Model}\label{sec:model}

In this section, we introduce the model of a quantum system interacting with an environment particle, undergoing a joint unitary evolution. This model constitutes the fundamental building block, which we  employ then to tackle the interaction of a quantum system with a complex environment leading to a non-equilibrium steady-state.
After having formulated the $1^{\rm st}$ law of thermodynamics for our model, we incorporate it into the framework of collision models, to compute numerically the open dynamics of the quantum system under scrutiny undergoing repeated interactions with identical auxiliary particles.

Let us introduce the model of a quantum system $S$ interacting with an auxiliary environment $A$. The two objects are governed by the respective Hamiltonians $H_S$ and $H_A$, and interact through $H_{\rm int}$. For the sake of simplicity, in the following we will consider the case where $S$ and $A$ are both qubit particles (see \Sref{sec:Analyticresults}). The theoretical framework, however, can easily be extended to generic discrete-level quantum systems, such as quantum harmonic oscillators. We define also the total Hamiltonian as $H_{SA} = H_S \otimes \mathbb{I}_A + \mathbb{I}_S \otimes H_A + H_{\rm int}$. In the paper, we will use the compact notation $\sigma_S \sigma_A$ to denote the tensor product of operators $\sigma_S \otimes \sigma_A$ acting on the Hilbert spaces of $S$ and $A$, and similarly for tensor products of system and environment density matrices. 
We always assume the initial joint state of system and environment to be factorized, $\rho_{SA} = \rho_S \rho_A$, without making further assumption on the quantum system's state. 

\subsection{Collision model}
\label{sec:collisionmodel}

Here we introduce the tool of collision models for open quantum system dynamics~\cite{CusumanoReviewEntropy,CiccarelloReview}.

In this approach, a quantum system $S$ interacts repeatedly, or ``collides'', with an environment made of many particles ${\{}A^{(n)}{\}} $. The system-particle interactions, lasting for the time $\tau$, occur within the Markovian regime so that the system collides each time with a different particle of the environment, which is then traced out. The initialisation state $\rho_{A}$ of the environment particles, before the single collision with the system, is always the same, while in our analysis the initial state of the system is considered as an arbitrary density operator. The dynamics of each interaction occurs exactly as described in \Sref{sec:1stlaw}. As an effect of repeated interactions, the state of the system changes, until reaching a steady-state after many collisions. The asymptotic state of the system may or may not be the same as $\rho_A$, depending on the specific choice of the interaction Hamiltonian $H_{\rm int}$. 

The system dynamics is computed in discrete steps at times $\{t = n \, \tau\}, \, n = 0,1,\dots$; at each step, a single interaction occurs between the system $S$ and the particle $A^{(n)}$ through the unitary operator 
\begin{equation}
U^{(n)}(\tau) = \exp \left( -\frac{ i }{\hbar}  H_{SA}^{(n)} \tau \right),
\label{eq:U_SA}    
\end{equation}
where the superscript $(n)$ indicates the interaction with $A^{(n)}$. For the state of the system, we will adopt the compact notation $\rho_S^{(n)} \equiv \rho_S (n \, \tau)$. Furthermore, whenever not needed explicitly,  we will omit the dependence on $\tau$ and just denote $U^{(n)} \equiv U^{(n)}(\tau)$.

The iteration procedure is carried out as follows. Starting from the system state $\rho_S^{(n-1)}$ at step $n-1$, and from the ``fresh'' environment particle $\rho_A^{(n)}$ initialized in the default state $\rho_A$, we have
\begin{eqnarray}
\rho_{SA}^{(n)} &=& U^{(n)} \big( \rho^{(n-1)}_S \rho_A \big) U^{(n)\,\dagger},
\label{eq:collisionmodel1}
\\
\rho_S^{(n)} &=& \Tr_A \left[ \rho_{SA}^{(n)} \right],
\label{eq:collisionmodel2}
\end{eqnarray}
and similarly for the post-interaction environment particle state. We assume that $H_{SA}^{(n)}$ is always the same type of Hamiltonian for every $A^{(n)}$, therefore from now on we will drop the superscript $(n)$ whenever focusing on a generic collision, and simply denote $H_{SA} \equiv H_{SA}^{(n)}$ and $U \equiv U^{(n)}$. Furthermore, from~\eref{eq:collisionmodel1} and \eref{eq:collisionmodel2}, one can infer that, in general, the asymptotic state $\rho^{\infty}_S$ satisfies $\rho^{\infty}_S = \Tr_A [U \big( \rho^{\infty}_S \rho_A \big) U^{\dagger}]$.  

We stress that this model is Markovian because the system interacts only once with each auxiliary environment particle, and never again; in practice, each environment particle is traced out after its interaction with the system. Furthermore, no interactions among environment particles are present, therefore the resulting system dynamics does not experience memory effects.

\subsection{First law of thermodynamics}
\label{sec:1stlaw}

In this section we introduce the thermodynamics of the two quantum systems $S$ and $A$ undergoing the interaction process effected by the unitary operator of \eref{eq:U_SA}. The treatment presented here is rather general and does not depend on the specific interaction model; nonetheless, it may be helpful to think of the process as one single collision between a quantum system and a generic environment particle $A$.

Let us assume the interaction between $S$ and $A$ is active only during a finite-time window of duration $\tau$, between the initial and final times $t_1$ and $t_2$, with $t_2=t_1 + \tau$. Thus, as introduced above, the joint system-environment state evolves unitarily through $\rho_{SA}(t_2) = U \rho_{SA}(t_1) U^{\dagger}$, from which we can readily compute the evolved marginal system and environment states $\rho_S(t_2) = \Tr_A [\rho_{SA}(t_2)]$ and $\rho_A(t_2) = \Tr_S [\rho_{SA}(t_2)]$. At any time $t$, the total internal energy of the compound $S+A$ system is generally given by $E_{SA} \equiv \Tr[H_{SA} \, \rho_{SA}(t)]$. However, given that at $t < t_1$ and $t > t_2$ the total Hamiltonian is just the sum of the two free local Hamiltonians, at such times the total internal energy splits into two local components:
\begin{eqnarray}
\label{eq:internalenergy}
E_{SA}(t) &=& \Tr\left[ (H_{S} + H_{A}) \rho_{SA}(t) \right] =
\nonumber
\\
&=& \Tr\left[ H_{S} \rho_{S}(t) \right] + \Tr\left[ H_{A} \rho_{A}(t) \right] =
\\
&\equiv& E_S(t) + E_A(t) \quad {\rm for} \quad t<t_1 \vee t>t_2. 
\nonumber
\end{eqnarray}
As a consequence, the total \emph{internal energy variation} between $t=t_1^-$ and $t=t_2^+$, due to the interaction process, can also be split into local contributions, 
\begin{eqnarray}
\label{eq:envariation}
\delta E_{SA}(\tau) &\equiv& E_{SA}(t_2) - E_{SA} (t_1) = \nonumber \\
&=& E_S(t_2) + E_A(t_2) - E_S(t_1) - E_A(t_1) \equiv
\\
&\equiv& \delta E_{S}(\tau) + \delta E_{A}(\tau). \nonumber
\end{eqnarray}
Notice that, even though no other system is considered beyond $S$ and $A$, $\delta E_{SA}(\tau)$ can be generally different from zero. The total energy variation is due exclusively to the action of the unitary operator $U$, and to the underlying interaction $H_{\rm int}$. It is not due e.g.~to heat flows to/from an additional heat reservoir. Therefore, we interpret $\delta E_{SA}(\tau)$ as \emph{non-energy-preserving work}, and~\eref{eq:envariation} represents an expression of the $1^{\rm st}$ \emph{law of thermodynamics} for the joint $S+A$ system, in the absence of heat flows.

We introduce now the special case of \emph{energy-preserving interactions}. This is the case when the interaction preserves the total energy during the process, that is, following~\eref{eq:envariation},
\begin{equation}
\label{eq:enconservation}
\delta E_{SA}(\tau) = 0 \quad \Longrightarrow \quad \delta E_{S}(\tau) = - \delta E_{A}(\tau),
\end{equation}
and it can be shown (see~\cite{DeChiaraNJP2018} and Appendix~\ref{sec:en-preserving-proof}) that a sufficient condition for this to happen is
\begin{equation}
\label{eq:enpreserving}
[H_{\rm int},H_S+H_A] = 0 \quad \Longleftrightarrow \quad [U(\tau),H_S+H_A] = 0 \quad \forall \tau.     
\end{equation}

\subsection{Why nonequilibrium collision models?}

Before delving into technical details, we would like to make a remark on the motivations behind the nonequilibrium approach studied here, using the tool of Kirkwood-Dirac quasiprobability distributions. The nonequilibrium approach brings us to evaluate not just the average values of the thermodynamic quantities of interest (which is provided by the first statistical moment of the corresponding quasiprobability distribution), but also higher statistical moments, and in particular the variance. This is necessary because quantum fluctuations play a relevant role, e.g.~for the following reasons:
\begin{itemize}
    \item 
    We are dealing with finite-dimensional quantum systems, regarding both the quantum system under scrutiny and the particles of the environment, whose dimension is comparable to that of the quantum system.
    \item 
    We take into account initial states containing quantum coherence with respect to the local Hamiltonian operators. In closed quantum systems evolving unitarily, the presence of coherence naturally causes oscillations between the energy eigenvectors.
    \item 
    The behaviour and role of statistical distributions of thermodynamic quantities in nonequilibrium quantum processes is still partly unexplored. In particular, the variance seems to play a special role in the characterization of nonequilibrium steady-states, whereby the average energy can be zero, but not the variance of the corresponding distribution (see e.g.~\cite{GherardiniTutorial} for a discussion on the role of work variance in the Kirkwood-Dirac quasiprobability setting, introduced in the next section). This implies that the energy of such a system is constantly fluctuating, even though this is not evident looking at the average alone. 
\end{itemize}

\section{Quasiprobability distributions}
\label{sec:KDQ}

Let us consider the Kirkwood-Dirac quasiprobability (KDQ) distribution of the stochastic internal energy variations in $S$ and $A$ for each collision, as well as of the non-energy-preserving work. Such distributions are built to guarantee that the average values over many realizations of the system's dynamics return the energy differences $\delta E_{S}^{(n)},\delta E_{A}^{(n)},\delta E_{SA}^{(n)}$ obeying the first law of thermodynamics for any $n$. The obtained quasiprobability distributions are equal to the ones provided by the two-point measurement (TPM) scheme~\cite{EspositoRMP2009} if the states of the system and the environment particles do not contain quantum coherence with respect to the local Hamiltonians $H_{S},H_{A}$.

We start from introducing the stochastic instances $u_S, u_A, u_{S+A}$ of the internal energy variations as well as of the non-energy-preserving work. They are given respectively by the differences of the eigenvalues (i.e., the energies) of the Hamiltonian operators $H_S, H_A, H_S+H_A$, where we introduce the spectral decompositions $H_S=\sum_{\ell}E^{S}_{\ell}\Pi^{S}_{\ell}$ and $H_A=\sum_{k}E^{A}_{k}\Pi^{A}_{k}$ with $\Pi^{L}_{k}\Pi^{L}_{j} = \Pi^{L}_{k}\delta_{k,j}$ ($L=S,A$). Notice that $H_S+H_A=\sum_{\ell,k}E^{S+A}_{\ell,k}\Pi^{S+A}_{\ell,k}$ where $E^{S+A}_{\ell,k} \equiv E^{S}_{\ell} + E^{A}_{k}$ and $\Pi^{S+A}_{\ell,k} \equiv \Pi^{S}_{\ell} \otimes \Pi^{A}_{k}$. In our analysis, the energies are evaluated at the beginning ($\ell=\ell_{\rm in}, k=k_{\rm in}$) and at the end ($\ell=\ell_{\rm fin}, k=k_{\rm fin}$) of each collision. Since the Hamiltonian operators are time-independent, one has that
\begin{eqnarray}
\label{eq:stochastic_instances}
&u_S (\ell_{\rm in},\ell_{\rm fin}) = E^{S}_{\ell_{\rm fin}} - E^{S}_{\ell_{\rm in}}, \label{eq:def_uS} 
\\
&u_A(k_{\rm in},k_{\rm fin}) = E^{A}_{k_{\rm fin}} - E^{A}_{k_{\rm in}},\label{eq:def_uA} 
\\
&u_{S+A}(k_{\rm in},\ell_{\rm in}, \ell_{\rm fin},k_{\rm fin}) = E^{S+A}_{\ell_{\rm fin},k_{\rm fin}} - E^{S+A}_{\ell_{\rm in},k_{\rm in}}\,.
\end{eqnarray}

Let us now define the KDQ distribution ${\rm P}_n(x)$ of $x = u_S, u_A, u_{S+A}$ for any collision $n$. Such a distribution reads as
\begin{equation}
\label{eq:distribution}
    {\rm P}_n(x) = \sum_{i_{\rm in},i_{\rm fin}} \mathfrak{q}_n\left( x(i_{\rm in},i_{\rm fin}) \right) \delta(x-x(i_{\rm in},i_{\rm fin}))
\end{equation}
where $\delta(\cdot)$ is the Dirac's delta function, $i=\ell,k$, and the quantity $\mathfrak{q}_n\left( x(i_{\rm in},i_{\rm fin}) \right)$ is the quasiprobability that describes the two-time probability of recording the value $x = x(i_{\rm in},i_{\rm fin})$ for $x = u_S, u_A, u_{S+A}$. By summing over all the possible initial and final energy values~\eref{eq:distribution} yields the complete KDQ distribution $P_n(x)$ for any arbitrary value of $x$. We recall that, when the state of the measured system (here, $\rho_S^{(n)}$ and $\rho_A$) does not commute with the measurement observables ($H_S$ and $H_A$), the quasiprobabilities $\mathfrak{q}_n$ could lose positivity, becoming complex numbers with negative real parts. 

In a coherent collision model \cite{RodriguesPRL2019,HammamNJP2022,HammamPRR2024}
\begin{equation}
\label{eq:ancilla}
    \rho_{A} = \rho_{A}^{\rm th} + \lambda \, \chi_A,
\end{equation}
where $\rho_{A}^{\rm th} \equiv e^{-\beta \, H_{A}}/{\rm Tr}[e^{-\beta \, H_{A}}]$, then $\chi_{A}$ is a Hermitian operator with no diagonal elements in the eigenbasis of $H_A$, and $\lambda$ is the control parameter tuning the magnitude of quantum coherence. Thus, in the Markovian regime, with an initial product state $\rho_{S}^{(n-1)}\rho_{A}$ before the $n$-th collision, the corresponding KDQs $\mathfrak{q}_n\left( x(i_{\rm in},i_{\rm fin}) \right)$ have the following expressions:
\begin{eqnarray}    
    &&\mathfrak{q}_n\left( u_S \right) = {\rm Tr}\left[ U^{\dagger} (\Pi^{S}_{\ell_{\rm fin}}\otimes\mathbb{I}_A) U (\Pi^{S}_{\ell_{\rm in}}\otimes\mathbb{I}_A) 
    \rho_{S}^{(n-1)} \rho_{A}\right]
    \label{eq:QP_uS}\\ &&\mathfrak{q}_n\left( u_A \right) = {\rm Tr}\left[ U^{\dagger} (\mathbb{I}_S \otimes \Pi^{A}_{k_{\rm fin}}) U (\mathbb{I}_S \otimes \Pi^{A}_{k_{\rm in}})     \rho_{S}^{(n-1)} \rho_{A}\right]
    \label{eq:QP_uA}\\
    &&\mathfrak{q}_n\left( u_{S+A} \right) = {\rm Tr}\left[ U^{\dagger} \Pi^{S+A}_{\ell_{\rm fin},k_{\rm fin}} U \, \Pi^{S+A}_{\ell_{\rm in},k_{\rm in}} \rho_{S}^{(n-1)} \rho_{A}\right],
    \label{eq:QP_uSplusA}
\end{eqnarray}
where we omitted indexes $(i_{\rm in},i_{\rm fin})$ in $u_S$, $u_A$ and $u_{S+A}$ for brevity of notation, as we will do from now onwards. The quasiprobabilities \eref{eq:QP_uS}-\eref{eq:QP_uSplusA} each sum to $1$ as they are built over proper density operators. Furthermore, they are built using the unitary operator $U$ governing the whole dynamics. Hence, the results provided by such quasiprobabilities hold in principle for any choice of parameters' values and thus even for an arbitrary value of $\tau$. This aspect is beneficial when investigating the distribution of stochastic energy changes between two consecutive collisions. The whole distribution is relevant in nonequilibrium regimes where fluctuations need to be taken into account.

\subsection{Averages}

Another hallmark of investigating nonequilibrium physics of collision models is the possibility to access the complete distribution of internal energy variations and non-energy-preserving work for any value of the parameters and collision time $\tau$. Let us start from the averages: from direct calculations, the expression of $\langle x^{(n)}\rangle$ for $x=u_S,u_A,u_{S+A}$ is:
\begin{eqnarray}
    &&\langle u_S^{(n)}\rangle = {\rm Tr}\left[\left( H_S \otimes \mathbb{I}_A  \right)\left( U\,\rho_{S}^{(n-1)} \rho_{A}\,U^{\dagger} - \rho_{S}^{(n-1)} \rho_{A} \right)\right], \label{eq:average_Us}
    \\
    &&\langle u_A^{(n)}\rangle = {\rm Tr}\left[\left( \mathbb{I}_S \otimes H_A  \right)\left( U\,\rho_{S}^{(n-1)} \rho_{A}\,U^{\dagger} - \rho_{S}^{(n-1)} \rho_{A} \right)\right], \label{eq:average_Ua}
    \\
    &&\langle u_{S+A}^{(n)}\rangle = {\rm Tr}\left[\left( H_S + H_A \right)\left( U\,\rho_{S}^{(n-1)} \rho_{A}\,U^{\dagger} - \rho_{S}^{(n-1)} \rho_{A} \right)\right], \label{eq:average_Us+a}
\end{eqnarray}
where $\langle u_S^{(n)}\rangle=\delta E_{S}^{(n)}$, $\langle u_A^{(n)}\rangle=\delta E_{A}^{(n)}$, and $\langle u_{S+A}^{(n)}\rangle=\delta E_{SA}^{(n)}$, as expected.

\subsection{Variances}

The analysis of nonequilibrium physics for collision models also allows to determine the expression of 
\begin{equation}\label{eq:variance_x}
    \left( \Delta x^{(n)} \right)^2 = \left\langle (x^{(n)})^{2}\right\rangle - \left\langle x^{(n)} \right\rangle^2
\end{equation}
for $x=u_S,u_A,u_{S+A}$, representing the variance of the internal energy variations of $S$ and $A$, and non-energy-preserving work respectively. 
In~\eref{eq:variance_x},
\begin{equation}
	\label{eq:stat_moment_alpha}
    \left\langle (x^{(n)})^{\alpha} \right\rangle = \int dx \, {\rm P}_n(x) x^{\alpha} = 
    \sum_{i_{\rm in},i_{\rm fin}}\mathfrak{q}_n\left( x(i_{\rm in},i_{\rm fin}) \right) x(i_{\rm in},i_{\rm fin})^{\alpha}
\end{equation}
with $\langle x^{(n)}\rangle$ given by \eref{eq:average_Us}-\eref{eq:average_Us+a}. 
For any collision, $(\Delta x^{(n)})^2$ embodies the fluctuations of $x^{(n)}$ and, in particular, the tendency to exchange energy by the system and the environment particles: the greater is the variance, the greater is the exchanged energy rate between the system and the environment. In the following, we will analyse variances in detail, as well as their consequences for the energetics of Markovian collision models, using an analytical case study with interacting qubits.

\section{Distribution of coherent work and incoherent heat}
\label{sec:weakcolmodel}

In this section, we assume the validity of the energy-preserving condition $[U,H_S+H_A]=0$, such that the non-energy-preserving work always vanishes. We also assume that the time $\tau$ between consecutive interactions is taken sufficiently small that we can apply the second-order Baker-Campbell-Hausdorff series expansion in $\tau$ for the unitary operator in~\eref{eq:U_SA}, which describes the dynamics of the system-particle compound. These assumptions lead us to the weakly coherent collision models introduced in~\cite{RodriguesPRL2019}. 
The weakly coherent character is due to the specific choice of the environment particles' preparation state 
\begin{equation}
\rho_{A} = \rho_{A}^{\rm th} + \widetilde{\lambda}\sqrt{\tau}\chi_A,
\end{equation}
where $\rho_{A}^{\rm th}$ denotes the thermal state at some inverse temperature $\beta$. Notice that $\widetilde{\lambda}$ is a free parameter similar to $\lambda$ in~\eref{eq:ancilla}, but it differs from it in that $\widetilde{\lambda}$ has units of $[{\rm time}]^{-1/2}$. This becomes necessary after introducing the scaling factor $\sqrt{\tau}$ that allows the derivation of a proper master equation 
built over the second-order series expansion of $U(\tau)$ in $\tau$~\cite{RodriguesPRL2019}.

As an effect of the energy-preserving condition, the average internal energy variation of either the system or the single environment particle can be split unambiguously in coherent work $\mathcal{W}$ and incoherent heat $Q$, as determined in \cite{RodriguesPRL2019}. Moreover, under these assumptions we can refer indiscriminately to the system or to each particle of the environment in order to get the quasiprobability distribution of internal energy variations, coherent work and incoherent heat. Thus, in this section, we will determine also the distribution of coherent work and incoherent heat, which satisfy the first law of thermodynamics for any individual collision $n$. For our convenience, these calculations are carried out from the point of view of the colliding environment particles.

Assuming that the collision time $\tau$ between consecutive interactions is taken sufficiently small entails that, at the generic $n^{\rm th}$ collision, the quantum system and the single environment particle correlate according to the following relation:
\begin{equation}
	\label{eq:collision_dynamics}
	\rho_{SA}^{(n)} 
    = \rho_{S}^{(n-1)}\rho_{A} - \frac{i\tau}{\hbar} \left[H_{SA},\rho_{S}^{(n-1)}\rho_{A}\right]  
    - \frac{\tau^2}{2 \, \hbar^2}\left[H_{SA},[H_{SA},\rho_{S}^{(n-1)}\rho_{A}]\right],
\end{equation}
where the initial state can be any arbitrary density matrix. Notice that the right-hand side in~\eref{eq:collision_dynamics} stems directly from the second-order Baker-Campbell-Hausdorff series expansion in $\tau$ of the unitary operator $U(\tau)$, in which the total Hamiltonian is implicitly expressed as $H_{SA} = H_S + H_A + H_{\rm int}/\sqrt{\tau}$~\cite{RodriguesPRL2019}. In accordance with the two-particle interaction model of \Sref{sec:model}, if we trace out the colliding particle, then the state of the system $S$ after the $n$-th collision is $\rho_S^{(n)}={\rm Tr}_{\rm A}[\rho_{SA}^{(n)}]$. We recall here that, under the assumption of a Markovian collision model, the colliding particle is assumed to no longer interact with the quantum system after the collision. The effect of repeated collisions is to induce a change in the state of the quantum system, until reaching a steady-state, which may even have some quantum coherence.

In the small-$\tau$ limit, the reduced dynamics of the system given by~\eref{eq:collision_dynamics} can be approximated by the master equation 
\begin{equation}
\label{eq:masterequation}
\dot{\rho}_{S}(t) = - \frac{i}{\hbar} [H_S + \widetilde{\lambda} \, G, \rho_S(t)] + \mathcal{D}[\rho_S(t)]
\end{equation}
that is valid for any time $t$~\cite{RodriguesPRL2019}. In~\eref{eq:masterequation},
\begin{equation}
\label{eq:coherentcorrection}
    G = {\rm Tr}_A \left[ H_{\rm int}(\mathbb{I}_S \otimes \chi_A) \right]
\end{equation}
represents an additional unitary contribution connected to the coherent part of each environment particle. Moreover,
\begin{equation}
\label{eq:Dissipator}
    \mathcal{D}[\rho_S(t)] = - \frac{1}{2 \,\hbar^2}
    {\rm Tr}_{A}\left[ H_{\rm int},[H_{\rm int},\rho_{S}(t) 
    \rho_{A}^{\rm th}] \right]
\end{equation}
plays the role of the usual (thermal) dissipator in the Lindblad form~\cite{breuer2002theory}, here dependent only on $\rho_{A}^{\rm th}$, the thermal part of $\rho_{A}$. The quantum dynamics generated by the master equation \eref{eq:masterequation} is non-unital (the fixed point of the corresponding map is not the identity) and admits a unique steady state, which is independent of the initial state of the system.

\subsection{First law of thermodynamics in the small-\texorpdfstring{$\tau$}{tau} limit}
\label{sec:FirstLawSmallTau}

Due to the energy-preserving condition $[U,H_S+H_A]=0$, the non-energy-preserving work is always equal to zero, and the internal energy variation $\delta E_{S}^{(n)}$ of the system after the $n$-th collision can be unambiguously attributed to energy flowing to or from the environment particle, whose internal energy variation is $\delta E_{A}^{(n)}$. Formally, this means that
\begin{eqnarray}
    \delta E_{S}^{(n)} &=& \Tr\left[(H_{S}\otimes\mathbb{I}_A)(\rho_{SA}^{(n)}-\rho_{S}^{(n-1)}\rho_{A})\right]= \label{eq:Delta_Es} 
    \\
    &=& -\Tr\left[(\mathbb{I}_S\otimes H_{A})(\rho_{SA}^{(n)} -\rho_{S}^{(n-1)}\rho_{A})\right] = -\delta E_{A}^{(n)} \label{eq:Delta_Ea}
\end{eqnarray}
where $\mathbb{I}_S$, $\mathbb{I}_A$ denote the identity operators on $S,A$ respectively.

As anticipated previously, the internal energy variation $\delta E_{S}^{(n)}$ can be decomposed in two distinct contributions that stem respectively from the off-diagonal and diagonal terms of the environment particle's state, with respect to $H_A$. The former contribution to $\delta E_{S}^{(n)}$ corresponds to the so-called {\it coherent work} $\mathcal{W}^{(n)}$ that reads as~\cite{RodriguesPRL2019}
\begin{eqnarray}
\mathcal{W}^{(n)} 
&&= \frac{i}{\hbar} \, \widetilde{\lambda} \,  \tau \, {\rm Tr}\Big[\left[G, H_S\right]\rho_S^{(n-1)}\Big] =
\label{eq:coh_work_system} 
\\
&&= - \frac{i}{\hbar} \, \sqrt{\tau} \, {\rm Tr} \Big[\left[G_A^{(n-1)}, H_A\right]\rho_A\Big] = 
\label{eq:coh_work_ancilla}
\\
&&= -\frac{i}{\hbar}\sqrt{\tau} \, {\rm Tr} \Big[ H_A \left[ \rho_A, G_A^{(n-1)} \right]\Big],
\nonumber
\end{eqnarray}
where $G_A^{(n-1)} = {\rm Tr}_S[H_{\rm int}(\rho_S^{(n-1)}\otimes\mathbb{I}_A)]$. Notice that the coherent work is present whenever $\widetilde{\lambda} \neq 0$. Instead, the other contribution to $\delta E_{S}^{(n)}$, which originates from the diagonal elements of $\rho_{A}$, is the {\it incoherent heat}~\cite{RodriguesPRL2019}
\begin{equation}
\label{eq:incoh_heat}
    Q^{(n)} = 
    {\rm Tr} \Big[ H_{A} \mathcal{D}_n[\rho_A^{\rm th}] \Big] 
\end{equation}
where 
\begin{equation}\label{eq:dissipator_rho_th_A}
    \mathcal{D}_n[\rho_A^{\rm th}] = \frac{\tau}{2 \hbar^2} {\rm Tr}_{S}\left[ H_{\rm int},[H_{\rm int},\rho_{S}^{(n-1)}\rho_{A}^{\rm th}]\right].
\end{equation}
In the limit of small $\tau$, the first law of thermodynamics reads as
\begin{equation}\label{eq:first_law}
    \mathcal{W}^{(n)} + Q^{(n)} = \delta E_{S}^{(n)} = -\delta E_{A}^{(n)}
\end{equation}
that is valid for any collision $n$. 

\subsection{Quasiprobability approach}
\label{sec:QP}

Having reviewed the weakly-coherent collision model, we provide the quasiprobability distributions of the thermodynamic quantities in this model, recalling that the non-energy-preserving work is equal to zero.

The quasiprobabilities associated to the internal energy variations of the system and the environment particles are given by~\eref{eq:QP_uS}-\eref{eq:QP_uA}, provided the unitary operator $U$ obeys the energy-preserving condition $[U,H_S+H_A]=0$ and $\tau$ is taken sufficiently small to warrant a second order expansion of $U$ in $\tau$.

We introduce the stochastic instances $w$ and $q$ of the coherent work and incoherent heat respectively. In this regard, using the energy-preserving condition, we determine them from addressing the environment particles, albeit they can be defined even by considering the quantum system only. We expand this aspect in Appendix~\ref{sec:SystemKDQ}.
In agreement with the first law of thermodynamics \eref{eq:first_law}, the stochastic instances $w$ and $q$ are both equal to $-u_A$, with $u_A$ defined by~\eref{eq:def_uA}. The KDQ $\mathfrak{q}_n( w(k_{\rm in},k_{\rm fin}) )$ and $\mathfrak{q}_n( q(k_{\rm in},k_{\rm fin}) )$ for the coherent work and incoherent heat, entering the distribution in~\eref{eq:distribution} with $x=w,q$ respectively, have thus the following expressions:
\begin{eqnarray}
&&\mathfrak{q}_n\left( w \right) = \widetilde{\lambda} \, {\rm Tr}\left[ U^{\dagger} (\mathbb{I}_S \otimes \Pi^{A}_{k_{\rm fin}}) U (\mathbb{I}_S \otimes \Pi^{A}_{k_{\rm in}}) \left(\rho_S^{(n-1)} \chi_A \right)\right],
\label{eq:QP_wc}
\\
&&\mathfrak{q}_n\left( q \right) = {\rm Tr}\left[ U^{\dagger} (\mathbb{I}_S \otimes \Pi^{A}_{k_{\rm fin}}) U (\mathbb{I}_S \otimes \Pi^{A}_{k_{\rm in}}) \left( \rho_S^{(n-1)}  \rho_{A}^{\rm th} \right) \right],
\label{eq:QP_qA}
\end{eqnarray}
with $U$ such that $[U,H_S+H_A]=0$, and $\tau$ small enough that $U(\tau)$ can be well approximated by its second order expansion in $\tau$.
It is worth noting that the quasiprobabilities $\mathfrak{q}_n( w )$ do not sum to $1$ but to $0$, since they are built over $\chi_A$, with ${\rm Tr}[\chi_A]=0$, representing the off-diagonal parts of the environment particle's state. Moreover, the fact that some of these quasiprobabilities $\mathfrak{q}_n( w )$ are different from zero entails that the distribution of the internal energy variations in both the system and the environment particle can have negative values as well as imaginary parts, which is a signature of non-classicality.

At the $n$-th collision, the average values $\langle w^{(n)}\rangle$ and $\langle q^{(n)}\rangle$ of the coherent work and incoherent heat are
\begin{eqnarray}
\langle w^{(n)}\rangle &=& \widetilde{\lambda}{\rm Tr}\left[\left( \mathbb{I}_S \otimes H_A  \right)\left( U\,\rho_{S}^{(n-1)} \chi_{A}\,U^{\dagger} - \rho_{S}^{(n-1)} \chi_{A} \right)\right],
\label{eq:average_Wc}
\\
\langle q^{(n)}\rangle &=& {\rm Tr}\left[\left( \mathbb{I}_S \otimes H_A  \right)\left( U\,\rho_{S}^{(n-1)} \rho_A^{\rm th}\,U^{\dagger} - \rho_{S}^{(n-1)} \rho_A^{\rm th} \right)\right],
\label{eq:average_Qa}
\end{eqnarray}
and it can be verified through direct calculation that, in the limit of small $\tau$, these estimates match the results previously obtained in \Sref{sec:FirstLawSmallTau}, $\langle w^{(n)}\rangle = \mathcal{W}^{(n)}$, see~\eref{eq:coh_work_system}, and $\langle q^{(n)}\rangle = Q^{(n)}$~\eref{eq:incoh_heat}, in agreement with the treatment introduced in~\cite{RodriguesPRL2019}.

\subsection{Operator approach}
\label{sec:OperatorApproach}

An alternative approach to study the fluctuations of coherent work is based on using a single system's observable. From~\eref{eq:average_Wc} we observe that the average coherent work can be written as the difference of two expectation values of two system's observables at the final and initial times:
\begin{equation}
    \langle w^{(n)}\rangle = \langle O_2\rangle_{S} -\langle O_1\rangle_{S},
\end{equation}
where we have defined the operators $O_1$ and $O_2$, acting on the generic system state $\rho_S$, as
\begin{eqnarray}
&O_1(\rho_S)& \, \equiv \widetilde{\lambda}\,{\rm Tr}_A \left[\left( \mathbb{I}_S \otimes H_A  \right) (\rho_S \otimes \chi_{A}) \right],
\\
&O_2(\rho_S)& \, \equiv \widetilde{\lambda} \, {\rm Tr}_A\left[U^\dagger\left( \mathbb{I}_S \otimes H_A  \right)U (\rho_S \otimes \chi_{A})\right].
\end{eqnarray}
The operator $O_1$ is by construction null, since we are working in the basis where $H_A$ is diagonal while $\chi_A$ has zero trace. Therefore, the statistics of $w^{(n)}$ for any $n$ can be reconstructed from the probability distribution of the system's observable $O_2$. To this end, let us write the spectral decomposition of $O_2 = \sum_i c^{O_2}_{i}\Pi^{O_2}_i$, where $c^{O_2}_{i}$ and $\Pi^{O_2}_i$ are the eigenvalues and corresponding eigenspace projectors of $O_2$. The probability of observing a certain eigenvalue of $O_2$ is then simply $p_i^{(n)} = {\rm Tr}[ \Pi^{O_2}_i \rho_{S}^{(n-1)} ]$ and hence the probability distribution for the coherent work is:
\begin{equation}
    \mathcal P(w^{(n)}) = \sum_i p_i^{(n)} \, \delta\left( w^{(n)} - c^{O_2}_{i} \right).
\end{equation}
The corresponding $k$-th moment can be calculated simply as:
\begin{equation}
    \left\langle ( w^{(n)} )^k \right\rangle = \sum_i p_i^{(n)} \left( c^{O_2}_{i} \right)^{k}.
\end{equation}
Notice that the probabilities $p_i^{(n)}$ are real, non-negative and sum to 1, since they are obtained as the expectation value of a positive semidefinite operator. 

\section{Results}
\label{sec:Analyticresults}

Let us deepen the meaning of the quasiprobabilities \eref{eq:QP_uS}-\eref{eq:QP_uSplusA} by providing analytical results for the case study of a concrete collision model. For this purpose, we consider that both the system and each environment particle are qubits whose (2-dimensional) Hilbert spaces are spanned by the computational basis $\{\vert 0 \rangle_L,\vert 1 \rangle_L\}$ with $L=S,A$. 
Thus, the local Hamiltonian operators read as
\begin{equation}
\label{eq:localH}
H_S = \frac{\hbar \, \omega_S}{2}\sigma_z^{S}, \quad H_A= \frac{\hbar \, \omega_A}{2}\sigma_z^{A},
\end{equation}
with $\hbar \omega_L>0$ denoting the local energy gaps and $\sigma_z^{L}=|0\rangle_{L}\!\langle 0| - |1\rangle_{L}\!\langle 1|$ the Pauli matrix $Z$ for the system and particle respectively ($L=S,A$). The two qubits interact through the following excitation-conserving Hamiltonian:
\begin{equation}
\label{eq:VSA}
H_{\rm int} = \frac{\hbar \, g}{2} (\sigma_{+}^{S}\sigma_{-}^{A} + \sigma_{-}^{S}\sigma_{+}^{A}), 
\end{equation}
where $g$ is the interaction strength in units of frequency, and $\sigma^{L}_{\pm} \equiv \frac{1}{2}(\sigma_x^{L} \pm i \, \sigma_y^{L})$ with $\sigma_x, \sigma_y$ Pauli matrices $X,Y$ ($L=S,A$). The interaction \eref{eq:VSA} is always excitation-preserving, namely $[U,n_S+n_A]=0$, with $n_L = \sigma^L_+ \,\sigma^L_-$. However, only in the \emph{resonant} case $\omega_S = \omega_A$ the energy-preserving condition $[U,H_S+H_A]=0$ is valid. Otherwise, for $\omega_S \neq \omega_A$ the quantum system and the environment particle exchange quanta with different energies, albeit preserving the total number of exchanged quanta.

Before each collision $n$, the state of all the environment particles is given by the density operator 
\begin{equation}
\label{eq:rho_A_qubits}
\rho_A = 
\left(
{ \begin{array}{cc}
\frac{ e^{-\beta \, \hbar \, \frac{\omega_A}{2} } }{ Z_A } &  \lambda \\
\lambda  & \frac{ e^{\, \beta \, \hbar \, \frac{\omega_A}{2} } }{ Z_A } \\ 
\end{array} } \right),
\end{equation}
for some value of the inverse temperature $\beta$ and of the quantum coherence's magnitude $\lambda$. In~\eref{eq:rho_A_qubits}, $Z_A \equiv e^{-\beta \, \hbar \, \omega_A/2} + e^{\beta \, \hbar \, \omega_A/2}$ is the partition function of $\rho_A$. With reference to the general expression of the environment particle state~\eref{eq:ancilla} in our case study $\chi_A = \sigma_x^A$. Notice that $\lambda$ is constrained such that ${\rm det}(\rho_A)\geq 0$, i.e., $|\lambda|^2 \leq 1/Z_A^2$, so as to ensure that $\rho_A$ is positive semi-definite. On the other hand, we parametrize the density operator of the system as
\begin{equation}
\rho_S^{(n)} = 
\left(
{ \begin{array}{cc}
\rho_{11}^{(n)}  & \mathfrak{Re}[\rho_{12}^{(n)}] + i \mathfrak{Im}[\rho_{12}^{(n)}] \\
\mathfrak{Re}[\rho_{12}^{(n)}] - i \mathfrak{Im}[\rho_{12}^{(n)}]  & 1 - \rho_{11}^{(n)} \\
\end{array} } \right),
\end{equation}
where the three coefficients $\rho_{11}^{(n)}$, $\mathfrak{Re}[\rho_{12}^{(n)}]$ (real part of $\rho_{12}^{(n)}$) and $\mathfrak{Im}[\rho_{12}^{(n)}]$ (imaginary part of $\rho_{12}^{(n)}$) are real numbers satisfying the constraints $\rho_{11}^{(n)} \in [0,1]$ and $\rho_{11}^{(n)} ( 1-\rho_{11}^{(n)} ) \geq \mathfrak{Re}[\rho_{12}^{(n)}]^2 + \mathfrak{Im}[\rho_{12}^{(n)}]^2$ that guarantee the positive semi-definiteness of $\rho_S^{(n)}$ for any collision $n$.

Using the definitions of the KDQs introduced in \Sref{sec:KDQ}, we have computed the analytic expressions of the quasiprobabilities for this qubit-qubit collisional model, which can be found in Appendix~\ref{sec:appC}.
In the following, instead, we are going to outline the main results for the loss of positivity of KDQs, the behaviour of averages and variances of the system energy variation and non-energy-preserving work, especially in the out-of-resonance regime, and of the coherent work and incoherent heat in the resonant regime. For simplicity of notation, the superscript $(\cdot)^{(n)}$, denoting the value taken at the $n$-th collision, will be dropped, and included explicitly only when necessary.

\subsection{Non-positivity of KDQs}

We are going to examine how the behaviour of the KDQ distributions can serve as a witness of non-classicality for quantum collision models in the Markovian regime. In particular, we are interested in searching for quantum traits in the energy statistics that originate from the non-commutativity between either the state of the system or the state of the environment particle with their respective Hamiltonians. We focus on the KDQ $\mathfrak{q}_n\left( u_S \right)$, associated to the internal energy variations within the system, and on the KDQ of the non-energy-preserving work, $\mathfrak{q}_n\left( u_{S+A}\right)$. Notice that the special case of energy-preserving interaction (at resonance) entails $\delta E_A = - \delta E_S$ and $\Delta u_A^2 = \Delta u_S^2$ (see \Sref{sec:1stlaw} and Appendix~\ref{sec:en-preserving-proof}). Hence, the same analysis would be equally valid if we considered the KDQ $\mathfrak{q}_n\left( u_A \right)$.

From the analytical expressions of $\mathfrak{q}_n\left( u_S\right)$ shown in Appendix~\ref{sec:appC}, one can readily see that all four values of the quasiprobability $\mathfrak{q}_n( u_S)$ comprise a ``base'' thermal component, which is real and only depends on the system's populations $\{ \rho_{11}, 1-\rho_{11} \}$. In addition to the thermal components, there are ``quantum correction'' terms that depend on the quantum coherences present in both the system's and environment particle's states:
\begin{eqnarray}
\mathfrak{q}_n ( u_S) &=& \mathfrak{q}^{\rm th}_n ( u_S) + \mathfrak{q}^{\rm q}_n( u_S),\\
\mathfrak{q}^{\rm q}_n ( u_S) &\propto& i \lambda \Big( \pm \mathfrak{Re}\left[\rho_{12}\right] - i\mathfrak{Im}\left[\rho_{12}\right] \Big) \sin(2 g \tau ).
\end{eqnarray}
It is worth stressing that such corrections appear if both the states of the quantum system and the environment have quantum coherence. In the context of the collision models studied here, this circumstance is not a peculiar limit case; indeed, if the system undergoes repeated interactions with particles with quantum coherence, its state will also naturally develop quantum coherence as an effect of the collisions.

Let us look at the \emph{non-positivity functionals}~\cite{LostaglioQuantum2023,GherardiniTutorial}
\begin{equation}
\label{eq:nonclassical}
\mathcal{N_{\rm q}}\big( {\rm P}_n(x) \big) \equiv -1 + \sum_{\ell_{\rm in},\ell_{\rm fin}} \Big\vert \mathfrak{q}_n \Big( x(\ell_{\rm in},\ell_{\rm fin})  \Big) \Big\vert,
\end{equation}
for $x = u_S, u_A, u_{S+A}$, which are greater than zero if some genuine quantum behaviour is at play.

\begin{figure}[th]
	\centering
	\begin{subfigure}[t]{0.315\textwidth}
		\centering
		\includegraphics[width=\textwidth]{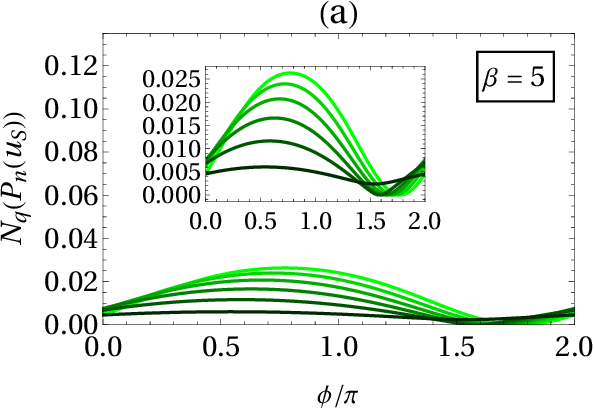}
	\end{subfigure}
	\hfill
	\begin{subfigure}[t]{0.285\textwidth}
		\centering
		\includegraphics[width=\textwidth]{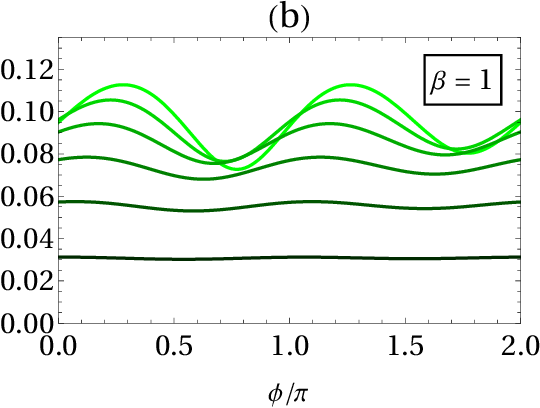}
	\end{subfigure}
	\hfill
	\begin{subfigure}[t]{0.37\textwidth}
		\centering
		\includegraphics[width=\textwidth]{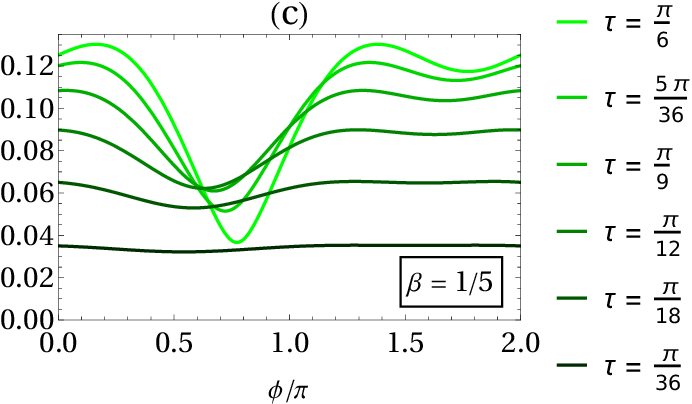}
	\end{subfigure}
	\caption{Non-positivity functional $\mathcal{N_{\rm q}}\left( {\rm P}_n(u_S) \right)$~\eref{eq:nonclassical} associated to the distribution of the system's internal energy variation, for the single collision (with $n$ set to $1$ without loss of generality). $\mathcal{N_{\rm q}}\left( {\rm P}_n(u_S) \right)$ is plotted for increasing collision time $\tau=\pi/36,\pi/18,\pi/12,\pi/9,5\pi/36,\pi/6$ and temperature $\beta=5,1,1/5$ as shown in panels (a), (b) and (c) respectively. These results are obtained in the non-resonant case with $\Delta = 3$, as a function of the phase $\phi$ of the quantum coherence term $\rho_{12}=r \, e^{i \, \phi}$ in the system's state. The values of the other model's parameters are: $\rho_{11}=1/4$, $\omega_A = 1$, $\omega_S = \omega_A+\Delta=4$, $g = 1$, $\hbar = 1$, $r=r_{\rm max}=\sqrt{\rho_{11}(1-\rho_{11})}=\sqrt{3}/4$, and $\lambda = \lambda_{\rm max} = 1/Z_A \simeq 0.082, \, 0.443, \, 0.498$ in panels (a), (b) and (c) respectively. Inset of panel (a): full-scale plot.}
	\label{fig:nonclassical_internal_energy}
\end{figure}

In Figure~\ref{fig:nonclassical_internal_energy}, we plot the behaviour of the non-positivity functional $\mathcal{N_{\rm q}}\left( {\rm P}_n(u_S) \right)$ built over the KDQ distribution of the system's internal energy variations. The plots are made as a function of the phase term $\phi \equiv {\rm arctg}\left( \mathfrak{Im}[\rho_{12}] / \mathfrak{Re}[\rho_{12}] \right)$, for some values of the collision time $\tau$ and inverse temperature $\beta$ of the environment particles' state. The values of the other relevant parameters in the model are fixed as detailed in caption of Figure~\ref{fig:nonclassical_internal_energy}.

We remark that the values of the collision time $\tau$ are chosen to be small enough so as not to substantially deviate from the original physical scenario of the weakly coherent collisional model described in \Sref{sec:weakcolmodel} and~\cite{RodriguesPRL2019}. Nonetheless, our approach based on quasiprobabilities does not require the small-$\tau$ limit in itself, since the quasiprobabilities of energy exchanges~\eref{eq:QP_uS}, \eref{eq:QP_uA}, \eref{eq:QP_uSplusA}, of coherent work~\eref{eq:QP_wc} and incoherent heat~\eref{eq:QP_qA} are built using the unitary $U$ of the entire system, with no approximations. While in principle one could always use arbitrary values of $\tau$ within our approach, we have chosen to keep $\tau$ reasonably small in order to maintain the similarity between the original definitions of coherent work~\eref{eq:coh_work_system} and incoherent heat~\eref{eq:incoh_heat}, and the corresponding non-equilibrium extensions based on quasiprobabilities, whenever needed. One can already see in Figure~\ref{fig:nonclassical_internal_energy} that, for the larger values of $\tau$, the curves start to depart from those obtained for smaller values of $\tau$. In the following, we present other calculations for values of $\tau$ up to $\pi$, where we observed a behaviour significantly different from the one presented here, embodying a dynamics in the strong coupling regime.

With regard to the parameter $\phi$, it corresponds to the phase of the quantum coherence entering the state of the quantum system. It is such that $\rho_{12} = r \, e^{i\, \phi}$ where $r$ is the modulus of $\rho_{12}$. In particular, keeping in mind the Bloch vector representation of the qubit state, $\phi=0$ implies that the coherence in the system is along the direction outlined by $\sigma_x$, whereas $\phi=\pi/2$ leads to the coherence along the direction corresponding to $\sigma_y$. In other words, the phase $\phi$ parametrizes a rotation of the system's Bloch vector on the equatorial plane of the Bloch sphere. Taking into account also that each environment particle is prepared in a state with real coherence terms (see~\eref{eq:rho_A_qubits} where $\lambda$ is real), one may effectively look at $\phi$ as the phase difference between the Bloch vectors associated to the initial states of the system and the environment particles, regardless of the moduli of coherences in these states.

From Figure~\ref{fig:nonclassical_internal_energy}, we observe that the non-positivity functional in general increases with the temperature and with the collision time $\tau$. These behaviours, however, are not strictly monotonic, since the functional presents a strong oscillatory dependence on the phase term $\phi$ for increasing collision time. In both the low and high temperature limits ($\beta=5$ and $\beta=1/5$ respectively), the non-positivity functional shows a dip for a specific value of the phase $\phi$ that is different depending on the temperature's value. Moreover, although it may seem counterintuitive that non-positivity increases with temperature, this arises from our constraint of taking the largest possible value for $r$ and $\lambda$, the magnitudes of quantum coherence entering the states of the quantum system and environment particles. In particular, from the plot, we can evince that the largest value of $\lambda$ increases with the temperature. We have verified that, if we choose a fixed value of $\lambda$ for the three inverse temperatures $\beta=5,1,1/5$, then the effect of having a decreasing value of the non-positivity functional with smaller temperature vanishes.

In numerical calculations we noticed that, as the amplitude $r$ of quantum coherence grows, the non-positivity functional $\mathcal{N_{\rm q}}\left( {\rm P}_n(u_S) \right)$ increases too. Moreover, we stress that, albeit the results in Figure~\ref{fig:nonclassical_internal_energy} are obtained in the non-resonant case with $\Delta = 3$, similar trends of $\mathcal{N_{\rm q}}\left( {\rm P}_n(u_S) \right)$ are observed setting $\Delta=0$, provided that quantum coherence is present in the states of the system and environment particles.

\begin{figure}[th]
	\centering
	\begin{subfigure}[t]{0.315\textwidth}
		\centering
		\includegraphics[width=\textwidth]{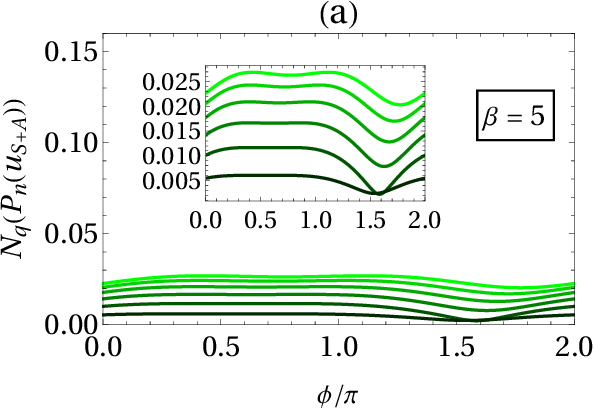}
	\end{subfigure}
	\hfill
	\begin{subfigure}[t]{0.285\textwidth}
		\centering
		\includegraphics[width=\textwidth]{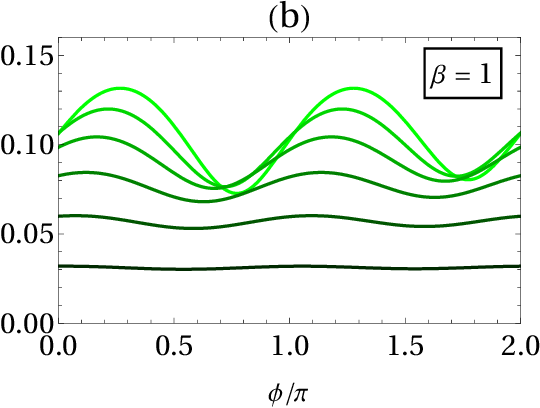}
	\end{subfigure}
	\hfill
	\begin{subfigure}[t]{0.37\textwidth}
		\centering
		\includegraphics[width=\textwidth]{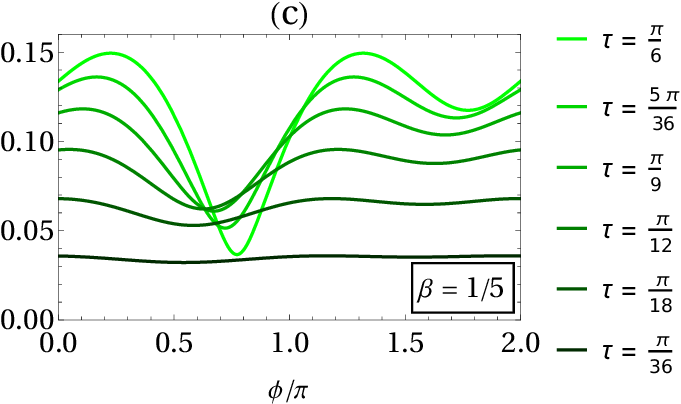}
	\end{subfigure}
	\caption{Non-positivity functional $\mathcal{N_{\rm q}}\left( {\rm P}_n(u_{S+A}) \right)$~\eref{eq:nonclassical} associated to the distribution of the non-energy-preserving work, for the single collision. $\mathcal{N_{\rm q}}\left( {\rm P}_n(u_{S+A}) \right)$ is plotted for increasing collision time $\tau=\pi/36,\pi/18,\pi/12,\pi/9,5\pi/36,\pi/6$ and temperature $\beta=5,1,1/5$ as shown in panels (a), (b) and (c) respectively. These results are obtained in the non-resonant case with $\Delta = 3$, as a function of the phase $\phi$ related to the quantum coherence $\rho_{12} = r e^{i \phi}$ of the system's state. The values of the other relevant model's parameters are: $\rho_{11}=1/4$, $\omega_A = 1$, $\omega_S = \omega_A+\Delta=4$, $g = 1$, $\hbar = 1$,
		$r=r_{\rm max}=\sqrt{\rho_{11}(1-\rho_{11})}=\sqrt{3}/4$, and $\lambda =  \lambda_{\rm max} = 1/Z_A \simeq 0.082, \, 0.443, \, 0.498$ in panels (a), (b) and (c) respectively. Inset of panel (a): full-scale plot.}
	\label{fig:nonclassical_non-energy-preserving-work}
\end{figure}

In Figure~\ref{fig:nonclassical_non-energy-preserving-work}, the non-positivity functional $\mathcal{N_{\rm q}}\left( {\rm P}_n(u_{S+A}) \right)$, built over the KDQ distribution of the non-energy-preserving work, is plotted as a function of $\phi$ and $\tau$. Similarly to the previous case, also $\mathcal{N_{\rm q}}\left( {\rm P}_n(u_{S+A}) \right)$ shows a monotonic increase with the magnitude $r$ of the coherence entering the quantum system's state, and it decreases for low temperatures ($\beta=5$, panel (a), in the figure) for the same reasons explained above. Also concerning the trend of the functional with $\phi$, for some collision time $\tau$, Figure~\ref{fig:nonclassical_internal_energy} and Figure~\ref{fig:nonclassical_non-energy-preserving-work} show evident similarities, although with some differences in the magnitude and the oscillatory behaviour.  

As alternatives to $\mathcal{N_{\rm q}}({\rm P}_n(x))$, one can consider two related formulations that quantify, respectively, the negativity of the real part and the non-reality of the quasiprobability distributions:
\begin{eqnarray}
\mathcal{N}_{{\rm Re}}\big(  {\rm P}_n(x)\big) &\equiv -1 + \sum_{\ell_{{\rm in}},\ell_{{\rm fin}} } \Big\vert \mathfrak{Re}[\mathfrak{q}_n ( x(\ell_{{\rm in}},\ell_{{\rm fin}}) )] \Big\vert, 
\label{eq:nonclassicalRE}
\\
\mathcal{N}_{{\rm Im}} \big(  {\rm P}_n(x) \big) &\equiv \sum_{\ell_{{\rm in}},\ell_{{\rm fin}} } \Big\vert \mathfrak{Im}[\mathfrak{q}_n ( x(\ell_{{\rm in}},\ell_{{\rm fin}}) )] \Big\vert \,,
\label{eq:nonclassicalIM}
\end{eqnarray}
Notice that $\mathcal{N}_{{\rm q}} \neq \mathcal{N}_{{\rm Re}} + \mathcal{N}_{{\rm Im}}$. 
Similarly to $\mathcal{N}_{{\rm q}}$, these other non-positivity functionals can also reveal the presence of quantum traits in energy distributions by taking values larger than zero.  
Using the abbreviated notation $\varphi \equiv g \tau$, we report the analytical expressions of $\mathcal{N}_{{\rm Re}}\left( {\rm P}_n(u_S) \right), \mathcal{N}_{{\rm Im}}\left( {\rm P}_n(u_S) \right)$ for the case study of the qubit-qubit collisional model with resonant interactions ($\Delta=0$), which read as:
\begin{eqnarray}
    \mathcal{N}_{{\rm Re}} = &-&1 + \displaystyle{\sum_{k=\pm 1}} \left\vert \sin(\varphi) \right\vert
    \bigg{\vert} \rho_k \frac{e^{\, k \, \frac{\beta \, \hbar \, \omega }{2}}}{Z_A} \sin(\varphi) + k \,  J_2 \, \cos(\varphi)\bigg{\vert} \nonumber 
    \\
    &+& \bigg{\vert} \frac{\rho_{k}}{2} \Big{(} 2 + k\frac{e^{-\frac{\beta \, \hbar \,\omega }{2}}}{Z_A} + \frac{e^{\, k\, \frac{\beta \, \hbar \, \omega}{2}}}{Z_A}\cos(2\varphi) \Big{)} - k\frac{J_2}{2} \sin (2\varphi)\bigg{\vert}, \label{eq:N_Re}
    \\
    \mathcal{N}_{{\rm Im}} = &2& \Big{\vert} J_1\sin(2\varphi) \Big{\vert},\label{eq:N_Im}
\end{eqnarray}
where $\rho_{+1} \equiv \rho_{11}$, $\rho_{-1} \equiv 1-\rho_{11}$,  $J_1 \equiv \lambda\,\mathfrak{Re}[\rho_{12}]$ and $J_2 \equiv 
\lambda\,\mathfrak{Im}[\rho_{12}]$. 
In~\eref{eq:N_Re} and \eref{eq:N_Im}, $\mathcal{N_{\rm Re}}\left( {\rm P}_n(u_S) \right)$ only depends on $\mathfrak{Im}[\rho_{12}]$, and conversely $\mathcal{N}_{{\rm Im}}\left( {\rm P}_n(u_S) \right)$ depends solely on $\mathfrak{Re}[\rho_{12}]$. Hence, if the coherence in the quantum system is such that $\mathfrak{Re}[\rho_{12}] \neq 0$, thus, with respect to $\sigma_x$ (the Bloch vector of $\rho_S$ has a non-zero component along the $x$-axis of the Bloch sphere), then this will lead to $\mathcal{N}_{{\rm Im}}\left( {\rm P}_n(u_S) \right) > 0$.  On the contrary, if the coherence in the quantum system is such that $\mathfrak{Im}[\rho_{12}] \neq 0$, thus with respect to $\sigma_y$ (Bloch vector along the $y$-axis), this will entail $\mathcal{N}_{{\rm Re}}\left( {\rm P}_n(u_S) \right) > 0$. Notice, however, that at the level of the quasiprobability $\mathfrak{q}_n(u_S)$, the correction terms due to quantum coherence start to be effective if the quantum coherence in~\eref{eq:nonclassicalRE} is strong enough to ``beat'' the thermal terms $\mathfrak{q}^{\rm th}_n( u_S )$, a condition that occurs when the temperature is low enough.

\subsection{Internal energy variations and non-energy-preserving work} 

\begin{figure}[th]
	\centering
	\begin{subfigure}[t]{0.495\textwidth}
		\centering
		\includegraphics[width=\textwidth]{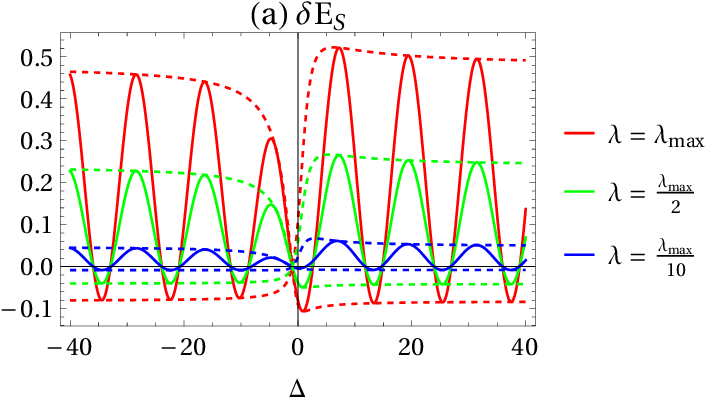}
	\end{subfigure}
	\hfill
	\begin{subfigure}[t]{0.495\textwidth}
		\centering
		\includegraphics[width=\textwidth]{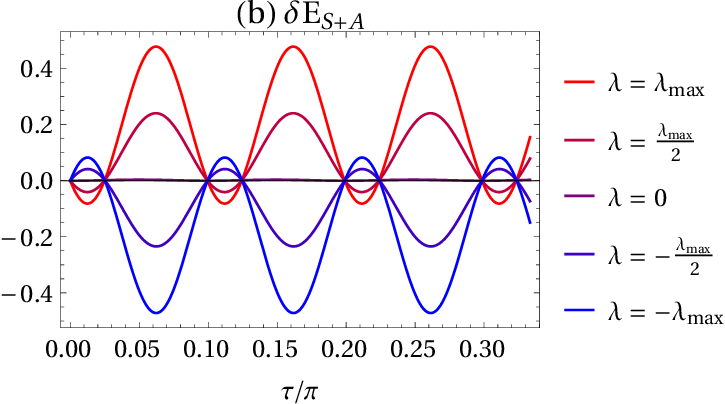}
	\end{subfigure}
	\caption{Average values of the system's internal energy variation and non-energy-preserving work in the out-of-resonance regime.
		Panel (a): Average $\delta E_S$ of the system's internal energy variation $u_S$ (solid lines) as a function of the detuning $\Delta = \omega_S - \omega_A$, for $\tau = \pi /6$ and various values of the quantum coherence $\lambda$ entering the environment particle's state. The dashed lines represent the upper and lower envelopes of $\delta E_S$, whose analytical expressions and asymptotic limits are given in the main text. Panel (b): Average $\delta E_{SA}$ of the non-energy-preserving work $u_{S+A}$ in the out-of-resonance regime with $\Delta = 20$, as a function of the collision time $\tau$, for various values of $\lambda$. The values of the other relevant model's parameters are: $\omega_A = 1$, $\omega_S = \omega_A+\Delta = 21$, $g = 1$, $\hbar = 1$, $\beta = 1$, $\rho_{11}=1/4$, 
		$r=r_{\rm max}=\sqrt{\rho_{11}(1-\rho_{11})}=\sqrt{3}/4$, $\phi = \pi/4$.}
	\label{fig:OutOfResonance}
\end{figure}

We now showcase relevant behaviours of averages and variances of both internal energy variations and non-energy-preserving work, for the qubit-qubit case study under scrutiny. By making use of the following auxiliary definitions:
\begin{equation}
\label{eq:functions}
\eqalign{
c_{\beta} &\equiv 1+e^{\, \beta \,\hbar \, \omega_A},
\cr
\tilde{\tau}(\Delta,\tau) &\equiv \tau \sqrt{4 \, g^2 +\Delta^2},  
\nonumber
\cr
a(\Delta) &\equiv \lambda \, c_{\beta} \, \mathfrak{Im}[\rho_{12}] \sqrt{4 \,g^2 +\Delta^2}, 
\cr
b(\Delta) &\equiv g \,(c_{\beta} \rho_{11} - 1) -\Delta \, \lambda \, c_{\beta}  \, \mathfrak{Re}[\rho_{12}],
\cr
\theta (\Delta) &\equiv \arctan \left( \frac{b(\Delta)}{a(\Delta)} \right),
}
\end{equation}
the average value $\delta E_{S}$ of the stochastic internal energy variation admits the following analytic expression:
\begin{equation}
\eqalign{	
\delta E_{S} &= \frac{2 \, \hbar \, g \, (\omega_A + \Delta)}
{c_{\beta}\, (4 g^2 + \Delta^2)}
\Big( g \,(1- c_{\beta} \, \rho_{11} ) + c_{\beta} \, \lambda \, \mathfrak{Re}[\rho_{12}] +
\cr
&- \sqrt{a^2(\Delta) + b^2(\Delta)} \sin \big( \tilde{\tau}(\Delta,\tau) - \theta(\Delta) \big) \Big),
}
\label{eq:deltaES}
\end{equation}
whose behaviour as a function of $\Delta$ is shown in Figure~\ref{fig:OutOfResonance}, panel (a), solid lines. As one can observe, $\delta E_{S}$ is an oscillatory function of $\Delta$, and the oscillation amplitude increases with the quantum coherence $\lambda$ entering the environment particle's state $\rho_A$. Moreover, the oscillations of $\Delta$ become stationary for large $|\Delta|$. This aspect is well captured by the upper and lower envelopes of $\delta E_{S}$ that are also included in Figure~\ref{fig:OutOfResonance} (a) as dashed lines. The analytical expression of the envelopes can be readily obtained from~\eref{eq:deltaES} by simply dropping the oscillating term $\sin \big( \tilde{\tau}(\Delta,\tau) - \theta(\Delta) \big)$.

In a similar fashion, we can determine the expression of the non-energy-preserving work $\delta E_{SA}$ that reads as
\begin{equation}
\label{eq:deltaESA}
\eqalign{
\delta E_{SA} &=
- \frac{4 \, \hbar \, g \, \Delta}{c_{\beta} \,(4 \, g^2 + \Delta^2)}
\sqrt{a^2(\Delta) + b^2(\Delta)} \times
\cr
&\times
\sin \bigg(\frac{\tilde{\tau}(\Delta,\tau)}{2}\bigg)
\cos \bigg( \frac{\tilde{\tau}(\Delta,\tau)}{2} - \theta(\Delta)  \bigg).
}
\end{equation}
The behaviour of $\delta E_{SA}$ as a function of $\Delta$ follows the same trend as $\delta E_{S}$. We have thus analysed $\delta E_{SA}$ as a function of the collision time $\tau$ and the quantum coherence $\lambda$. Panel (b) of Figure~\ref{fig:OutOfResonance} shows these behaviours in the extreme out-of-resonance regime, $\Delta \gg 1$. It is evident that $\delta E_{SA}$ has a periodic behaviour as a function of $\tau$, as expected due to the unitary character of the single collision dynamics. Furthermore, just as for $\delta E_{S}$, the amplitude of the oscillations increases with $\lambda$. Moreover, the sign of $\delta E_{SA}$ follows the sign of $\lambda$. All these features are well captured by a simplified expression of $\delta E_{SA}$ in the extreme out-of-resonance limit, obtained from~\eref{eq:functions} and \eref{eq:deltaESA} by taking the limit $\Delta \to \infty$ and using the parametrization for the quantum coherence $\rho_{12} = r \, e^{i \, \phi}$:
\begin{equation}\label{eq:deltaESA_simplified}
\delta E_{SA} \simeq  4 \, \hbar \, g \, \lambda \, r 
\sin \bigg( \frac{\Delta \, \tau}{2} \bigg)
\sin \bigg( \frac{\Delta \, \tau}{2} - \phi \bigg).
\end{equation}
We remark that the approximate expression above is the same whether we take the limit $\Delta \to +\infty$ or $\Delta \to -\infty$. Equation~\eref{eq:deltaESA_simplified} makes evident the important role played by the initial quantum coherence in both the amplitude and phase of the oscillations presented by $\delta E_{SA}$.

\begin{figure}[th]
	\centering
	\begin{subfigure}[t]{0.31\textwidth}
		\centering
		\includegraphics[width=\textwidth]{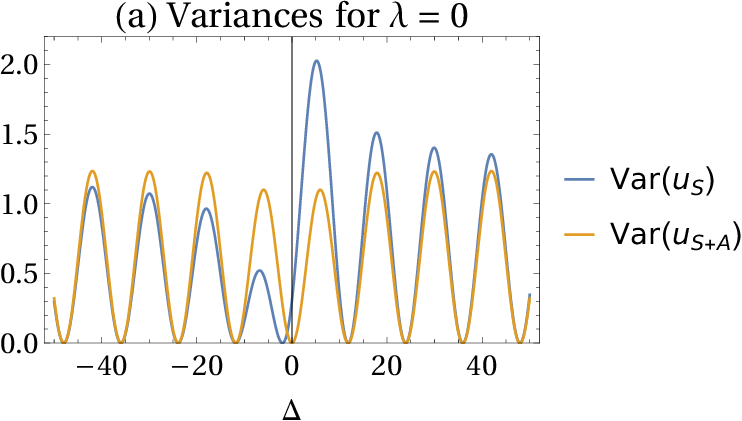}
	\end{subfigure}
	\hfill
	\begin{subfigure}[t]{0.32\textwidth}
		\centering
		\includegraphics[width=\textwidth]{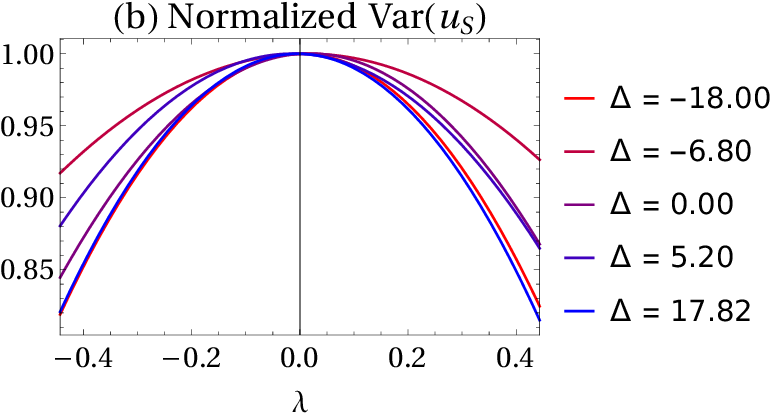}
	\end{subfigure}
	\hfill
	\begin{subfigure}[t]{0.32\textwidth}
		\centering
		\includegraphics[width=\textwidth]{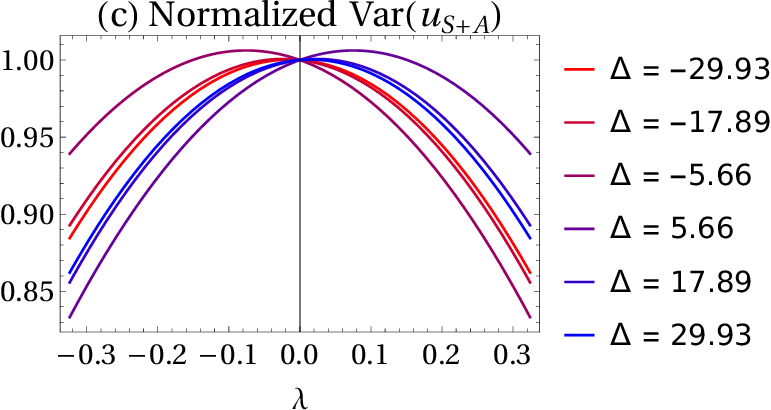}
	\end{subfigure}
	\caption{Variance $\Delta u_S^2$ of the system's internal energy variation, and variance $\Delta u_{S+A}^2$ of the non-energy-preserving work, as a function of the quantum coherence $\lambda$ and detuning $\Delta = \omega_S - \omega_A$. Panel (a): Variances versus $\Delta$ in the case with $\lambda =0$. Panels (b) and (c): $\Delta u_S^2$ and $\Delta u_{S+A}^2$, respectively, as a function of $\lambda$, for some values of $\Delta$ corresponding to the local maxima of the curve in panel (a). Variances are normalized with respect to the corresponding value for $\lambda = 0$. The values of the other relevant model's parameters are: $\rho_{11}=1/4$, $\omega_A = 1$, $\omega_S = \omega_A+\Delta$, $g = 1$, $\hbar = 1$, $\tau = \pi /6$, $r=r_{\rm max}=\sqrt{\rho_{11}(1-\rho_{11})}=\sqrt{3}/4$, $\phi = \pi/4$, $\beta = 1$.}
	\label{fig:Variance}
\end{figure}

In Figure~\ref{fig:Variance} we plot the behaviour of variances $\Delta u_S^2$ and $\Delta u_{S+A}^2$, respectively, for the system's internal energy variation and the non-energy-preserving work, as a function of $\Delta$ and $\lambda$. As shown in panel (a) setting $\lambda=0$, the trend with respect to $\Delta$ of both $\Delta u_S^2$ and $\Delta u_{S+A}^2$ is oscillatory, similarly to the average values $\delta E_S$, $\delta E_{SA}$. Then, we observe that the frequencies of the oscillations of $\Delta u_S^2$ and $\Delta u_{S+A}^2$ become stationary and coincide for large $\Delta$. The same oscillatory features of the variances is maintained even if the initial state of each environment particle is coherent with $\lambda \neq 0$. However, the largest value of the variance $\Delta u_S^2$ is for $\lambda=0$, and quadratically decreases as $|\lambda|$ increases, as plotted in Figure~\ref{fig:Variance}(b). This finding is in agreement with the evidence discussed in the literature that the presence of quantum coherence in the initial state of a thermodynamic transformation can entail a decrease in {\it local} energy fluctuations~\cite{hernandez2024projective,GherardiniTutorial,hernandez2024Interfero}. This effect is partially hidden in the variance of the non-energy-preserving work, which indeed accounts for fluctuations of both system's and environment particle's internal energy variations; see Figure~\ref{fig:Variance}, panel (c). However, a decreasing trend in $\Delta u_{S+A}^2$ similar to the one of $\Delta u_S^2$ is recovered for either large $|\lambda|$ or large $\Delta$  (extreme out-of-resonance regime).

\subsection{Coherent work and incoherent heat}

Let us now single out how the quasiprobability approach and the operator approach (OA) work differently when characterizing the coherent work; similar findings hold when characterizing the incoherent heat. Once more, we recall that this analysis finds proper application in the resonant regime of $\Delta=0$, and $\tau$ sufficiently small.

First of all, we provide the analytical expressions of the average and variance associated to the stochastic instance $w$ of the coherent work at the single collision, i.e.,
\begin{eqnarray}
    \langle w \rangle &=& \mathcal{W} = - \hbar \, \omega \, J_2 \sin(2 \varphi),
    \label{eq:averageWc}
    \\
    \Delta w^2 &=& - (\hbar \, \omega)^2 \sin(2 \varphi) \left( J_2^2 \sin(2\varphi) + i J_1 \right),
    \label{eq:VarianceWc}
\end{eqnarray}
where, we recall, $J_1 = \lambda \,\mathfrak{Re}[\rho_{12}]$ and $J_2 = \lambda \,\mathfrak{Im}[\rho_{12}]$ (see also Appendix~\ref{sec:appC}).
It is worth recalling that the distribution of coherent work is ill-defined, as the corresponding quasiprobabilities $\mathfrak{q}_n( w )$ sum to zero, being built over $\chi_A$ which is not a proper density matrix operator. However, the following symmetries are obeyed: $\mathfrak{Re}[\Delta w^2] = -\mathcal{W}^2$, and $\mathfrak{Im}[\Delta w^2] = \langle w^2 \rangle = -(\hbar \, \omega)^2 J_1 \sin(2 \varphi)$. 

\begin{figure}[th]
    \centering
    \includegraphics[width=0.6\textwidth]{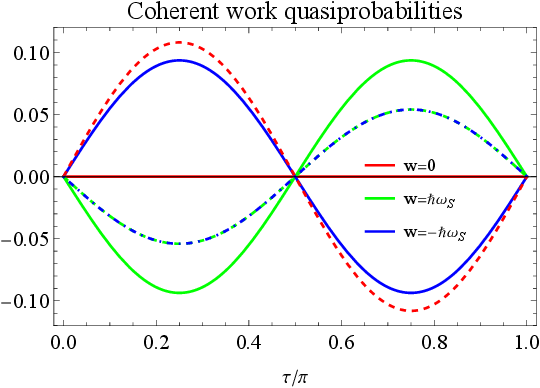}
    \caption{Quasiprobabilities of coherent work as a function of the collision time $\tau$, at resonance, displayed for each of the different values that $w$ can assume: $w=0$ (red), $w = \hbar \, \omega_S$ (green) and $w = - \hbar \, \omega_S$ (blue). Solid (respectively dashed) lines refer to the real (resp.~imaginary) part of the quasiprobabilities.
    The values of the other relevant model's parameters are: $\omega_S = \omega_A = 1$, $g = 1$, $\hbar = 1$, $\beta = 1/10$, $\rho_{11}=1/4$, 
    $r=r_{\rm max}=\sqrt{\rho_{11}(1-\rho_{11})}=\sqrt{3}/4$, $\phi = \pi/3$, $\lambda =  \lambda_{\rm max} = 1/Z_A \simeq 0.499$.}
    \label{fig:qCoherentWork}
\end{figure}

In Figure~\ref{fig:qCoherentWork} we plot the real (solid lines) and imaginary parts (dashed lines) of KDQs of coherent work as a function of the collision time $\tau$, for the largest possible value of $\lambda$. It can be observed that all functions oscillate as a function of $\tau$ and that the real parts of the quasiprobabilities become negative, with alternating signs. Coherent work and relative fluctuations vanish when either the state of the quantum system or that of the environment particle loses quantum coherence. 

\begin{figure}[th]
    \centering
    \includegraphics[width=0.6\textwidth]{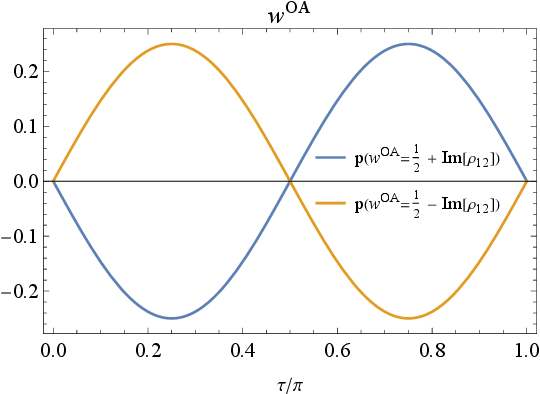}
    \caption{Coherent work at resonance, computed through the operator approach (OA), as a function of $\tau$. Here, the two stochastic values that $w$ can assume vary in time according to~\eref{eq:wOA}, while their probabilities remain fixed, as reported in the figure. The same parameters' values of Figure~\ref{fig:qCoherentWork} are used.
    }
    \label{fig:pCoherentWorkOA}
\end{figure}

In \Sref{sec:OperatorApproach} we introduced the operator approach as an alternative method to estimate the average value of the stochastic coherent work $w$, as well as the higher statistical moments of the corresponding distribution. To apply this method to our case study, we need to diagonalize the operator $O_2$, obtaining its eigenvalues $\{ c_i^{O_2} \}$ and eigenprojectors $\{ \Pi_i^{O_2} \}$, from which we compute the probabilities $p_i = {\rm Tr}[\rho_S \Pi_i^{O_2} ]$ given the input state $\rho_S$. Then, we obtain the desired result as $\langle w \rangle = \sum_i p_i c^{O_2}_{i}$. The specific definition of $O_2$ makes that its eigenvectors are constant in time, and so are the probabilities $\{ p_i \}$, which assume the two values $p_{\pm} = 1/2 \pm \mathfrak{Im}[\rho_{12}]$. Conversely, the eigenvalues $\{ c_i^{O_2} \}$ depend on the collision time $\tau$ and assume the meaning of the particular values $w$ can assume for the operator approach. Specifically, they read as 
\begin{equation}
    w^{OA}_{\pm}(\tau) \equiv c_{\pm}^{OA}(\tau) = \mp \frac{1}{2} \hbar \, \omega \, \lambda \sin (2 g \, \tau) \, ,
    \label{eq:wOA}
\end{equation}
and their behaviour in time is displayed in Figure~\ref{fig:pCoherentWorkOA}.

Quasiprobabilities and the operator approach to estimate the statistical moments of coherent work provide similar results, albeit they are complementary. In fact, considering the average of $w$, the use of quasiprobabilities leads to $\langle w \rangle = \sum_i \mathfrak{q}(w_i) w_i$, while with the operator approach one has $\langle w \rangle = \sum_i p_i w^{O_2}_{i}$. The two relations are perfectly equivalent, so that the (unperturbed) average value of the stochastic coherent work $w$ they return is the same. However, in the former case, the quasiprobabilities vary in time and the values assumed by the stochastic work are fixed, while in the latter case the opposite occurs. This discrepancy is due to the different ways the coherent work can be experimentally measured: indirectly according to KDQ and directly employing the operator approach.

With regard to the incoherent heat, instead, the average and variance of the corresponding quasiprobability distribution read respectively as
\begin{eqnarray}
\label{eq:averageQinc}
    \langle q_A \rangle &=& Q = \hbar \, \omega \sin^2(\varphi) \left( \frac{e^{-\beta \, \hbar \, \omega/2}}{Z} - \rho_{11} \right) 
    \\
    \Delta q_A^2 &=& (\hbar \, \omega)^2 \sin^2(\varphi)\left( \frac{e^{-\beta \, \hbar \, \omega /2} + \sinh(\frac{\beta \, \hbar \, \omega}{2}) \, \rho_{11} }{Z} \right) -  Q^2 \,,
\end{eqnarray}
where $Z = e^{\, \beta \, \hbar \,\omega /2} + e^{-\beta \, \hbar \, \omega /2}$. As expected, neither the average nor the variance exhibit any dependence on quantum coherence of the system's and environment particle's states.
Combining~\eref{eq:deltaES}, \eref{eq:averageWc} and~\eref{eq:averageQinc}, we can easily verify the validity of the 1$^{\rm st}$ law of thermodynamics for any collision in the small-$\tau$ limit:
\begin{equation}
    \mathcal{W} + Q = \delta E_{S} = -\delta E_{A}, 
\end{equation}
as discussed also in Appendix~\ref{sec:AppAveVar}.

\section{From theory to experimental tests}
\label{sec:proposal}

Collision models were introduced mainly as a tool to compute open quantum system dynamics, especially through numerical methods, in regimes not easily treatable with master equations, for instance when quantum correlations between system and environment, or quantum non-Markovianity play a significant role. As such, they are not usually meant to be taken literally, rather as effective computational tools. In spite of this, a variety of experimental scenarios can be considered to which the collision model explored here can be applied. Such setups can indeed constitute suitable test beds for the model, as well as for the thermodynamics of the physical processes described by the model. A requirement of particular relevance for our purposes is a sufficient degree of control of the quantum systems involved, namely the capacity of preparing their quantum states with a degree of quantum coherence. In fact, we recall that the quantum effects reported in \Sref{sec:Analyticresults} always depend on the product of system's and environment's coherence terms, thus requiring both to be present. The experimental setup on which we mainly focus our discussion is {\it microwave photonics with superconducting quantum circuits}. However, before introducing it, we mention few other platforms which could also constitute potential test beds.

In cavity QED systems~\cite{Haroche2007} a single photon in the cavity can represent the quantum system of interest, whereas atoms passing through the cavity, interacting sequentially with the photon, play the role of the environment particles of the collision model. This scenario probably represents an ideal test bed, because of the high degree of control in the preparation of both the photons' and the incoming atoms' quantum states.

Collision models proved to be especially useful to explore non-Markovianity of quantum dynamics, and this context has been investigated experimentally in all-optical setups such as~\cite{Bernardes2015}. Furthermore, in \cite{MaffeiPRR2021} it has been suggested to employ a waveguide quantum electrodynamics setup to probe non-classical light fields through thermodynamic witnesses. In \cite{MaffeiPRR2021}, a two-level system, embodied by an atom sitting in a light waveguide, is used to probe the statistics of the incoming light. It was found that the statistics of work extraction admits a bound which is satisfied if the incoming light is a coherent beam (with possible classical traits), and violated if the light is sent in single-photon pulses. This scenario could be a test bed for our theory because the ability to prepare single photon states would also allow for the preparation of superpositions of energy eigenstates, i.e., superpositions of different photon number states.

\begin{figure}[th]
    \centering
    \begin{subfigure}[t]{0.495\textwidth}
        \centering
        \includegraphics[width=\textwidth]{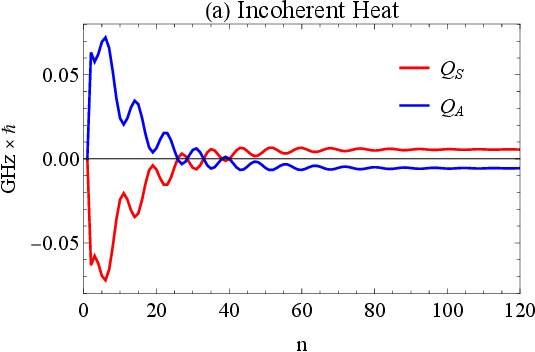}
    \end{subfigure}
    \hfill
    \begin{subfigure}[t]{0.495\textwidth}
        \centering
        \includegraphics[width=\textwidth]{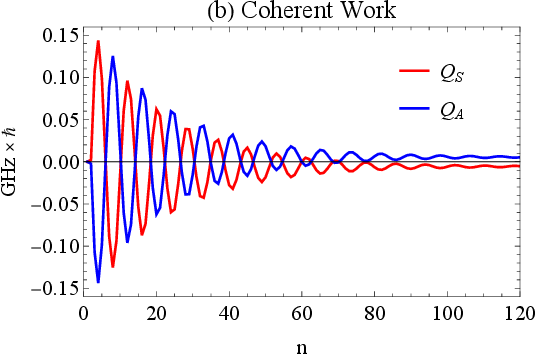}
    \end{subfigure}
    \caption{Incoherent heat and coherent work. In panel (a) and (b) we show the evolution of incoherent heat and coherent work, respectively, as a function of discretized time steps. The plots are obtained through numerical simulation by iterating the collision model~\eref{eq:collisionmodel1}-\eref{eq:collisionmodel2}, working at resonance ($\Delta=0$). In the simulations, the physical parameters are chosen according to the experimental scenario of microwave photonics with superconducting quantum circuits, i.e., $\omega_S = \omega_A = 5.7$ GHz, $g = 2/5 \, \omega_S$,  $\tau \simeq 135$ ps, $\beta = 0.4 \times 10^{-9} J^{-1}/\hbar$.
    Regarding state initialization instead, we chose:  $\rho_{11}=1/4$, 
    $r=r_{\rm max}=\sqrt{\rho_{11}(1-\rho_{11})}=\sqrt{3}/4$, $\phi = \pi/4$, $\lambda = \lambda_{\rm max}/2 \simeq 
    0.125$.}
    \label{fig:coherentwork}
\end{figure}

Finally, let us discuss a possible implementation of our model with the experimental platform of microwave photonics with superconducting quantum circuits~\cite{GU20171}, on which we based the estimates presented in the rest of this section. The working principle of superconducting quantum circuits is the tunnelling of Cooper pairs between small superconducting islands, leading to circuits based on Josephson junctions. These systems have shown a discretized energy structure, which allow the implementation of artificial atoms with a non-trivial spectrum, due to anharmonicities in the potentials that can be realized. Two main physical parameters determining the behaviour of the devices are the electrostatic Coulomb energy (or single-electron charging energy), and the Josephson energy. The ratio between these two energies defines different working regimes of the device. 

As a consequence, different types  of qubits can be realized, with the most important types being charge qubits, flux qubits and phase qubits. Since the energy scales involved are typically in the microwave range, a variety of physical scenarios have been explored where a superconducting quantum circuit is coupled to a microwave resonator. This allows for the exploration of qubit-photon interacting systems, similar to the ones studied in cavity QED in the optical regime and employing real atoms. In a qubit-photon coupled system, the Jaynes-Cummings Hamiltonian can be reached in a parameter regime where the rotating wave approximation is valid, that is $(\omega_S + \omega_A) \gg \{g, \Delta \}$, that is, the coupling is below the threshold of ultra-strong-coupling regime, and the qubit and resonator are not too far out of resonance. With reference to~\cite{GU20171}, we choose the physical parameters as  $\omega_S = \omega_A = 5.7$ GHz, $g = 2/5 \, \omega_S$, $\tau \simeq 135$ ps, $\beta = 0.4 \times 10^{-9} J^{-1}/\hbar$.

With this choice of parameters, we simulate the dynamics obtained by iterating the collision model, according to~\eref{eq:collisionmodel1} and~\eref{eq:collisionmodel2}. As an example, we show in Figure~\ref{fig:coherentwork} the discrete-time evolution of incoherent heat and coherent work at resonance ($\Delta = 0$), computed with the quasiprobability approach~\eref{eq:average_Wc} and \eref{eq:average_Qa}. As it is evident in the plots, we have both $\langle q_S \rangle = -\langle q_A \rangle$ and $\langle w_S \rangle = - \langle w_A \rangle$, implying $\delta E_S = -\delta E_A$, which verifies global energy preservation, as required by the condition of resonant interaction. It is still worth emphasizing that a non-zero coherent work is conditional on both the system and the environment particles being prepared in states with a degree of quantum coherence (cf.~the chosen parameters in caption of Figure~\ref{fig:coherentwork}), as it also emerges from the analytical expression in~\eref{eq:averageWc}.   

\section{Conclusions}

In this paper we determine the probability distribution of relevant thermodynamic quantities in a Markovian collision model. In particular, we take into account the possibility that the state of both the quantum system and each environment particle, colliding with it, can have quantum coherence with respect to the corresponding local Hamiltonians. Having states with quantum coherence is one of the mechanisms that allow to achieve a nonthermal, nonequilibrium steady-states~\cite{hernandezPRXQuantum2022,YadaPRL2022} in Markovian collisional models.

The derivation of probability distributions at multiple times under the presence of quantum coherence and correlations is achieved by resorting to a quasiprobability distribution, the Kirkwood-Dirac distribution in our paper~\cite{GherardiniTutorial}. In this way, we are able to determine the distribution of internal energy variations both for the quantum system and each particle of the Markovian environment under generic conditions. These conditions account for the non-energy-preserving scenario (achieved in this paper through a non-resonant interaction, namely $\Delta \neq 0$), and even for an arbitrary long collision time $\tau$. Moreover, our results apply to both the single collision and a sequence of them leading to the system's steady-state. Our approach offers a new perspective in the analysis of collisional models in a non-equilibrium context, starting with the calculation of the variances of exchanged energy distributions. Indeed, for finite-dimensional systems, variances are as important as the averages of the distributions, because they provide the rate at which energy is absorbed and emitted (and vice versa) by each subsystem.

As discussed in \Sref{sec:proposal}, our results can find application to many quantum platforms ranging from cavity QED to waveguide QED. Among these setups, microwave photonics with superconducting quantum circuits seems the most promising, since the variety of operational regimes permitted by the platform allows for the implementation of a Jaynes-Cummings interaction between a superconducting qubit and microwave photons. In this regime, the conditions that most closely match our model could be realized, and the coherent work, incoherent heat and their fluctuation features could be measured. 

\ack

The authors acknowledge enlightening discussions with Kenza Hammam, Alessandra Colla, Salvatore Lorenzo, Mauro Paternostro, Andrea Smirne, and Bassano Vacchini.
M.P. and S.G.~acknowledge financial support from the PRIN project 2022FEXLYB Quantum Reservoir Computing (QuReCo). S.G.~thanks the PNRR MUR project PE0000023-NQSTI funded by the European Union---Next Generation EU. G.D.C.~and S.G.~acknowledge support from the Royal Society Project IES\textbackslash R3\textbackslash 223086 ``Dissipation-based quantum inference for out-of-equilibrium quantum many-body systems''.

\section*{Data Availability Statement}

The code implementing the computations for the figures of the paper is available at~\cite{MarcoGithub}. 

\appendix

\section{Condition for energy-preserving interaction}\label{sec:en-preserving-proof}

In this Appendix we show, for clarity, that the condition for having energy-preserving interactions~\eref{eq:enconservation} holds if $[H_{\rm int},H_S+H_A] = 0$ (see also~\cite{DeChiaraNJP2018}).

Let us consider the total energy variation between times $t=0^-$ and $t=\tau^+$, that is, before and after the interaction comes into effect:
\begin{equation}
\eqalign{	
    \delta E_{SA}(\tau) &= \Tr \left[ (H_S + H_A) \rho_{SA}(t_2) \right] - \Tr \left[ (H_S + H_A) \rho_{SA}(t_1) \right] =
    \\
    &= \Tr \left[ (H_S + H_A)  U \rho_{SA}(t_1) U^{\dagger} \right] 
    - \Tr\left[ (H_S + H_A)  \rho_{SA}(t_1) \right].
}
\end{equation}
Let us use the ciclicity of the trace in the last formula, and recall that $U = \exp (- i (H_S+H_A + H_{\rm int})\tau/\hbar)$. We thus obtain that, if $[H_{\rm int},H_S+H_A]=0$, then $U = \exp (- i (H_S+H_A)\tau/\hbar) \exp(- i H_{\rm int}\tau/\hbar)$, and
\begin{equation}
U^{\dagger} (H_S + H_A)  U = H_S + H_A,    
\end{equation}
which leads straightforwardly to $\delta E_{SA} = 0$, therefore the total energy is conserved by the process.

\section{Stochastic coherent work and incoherent heat, and their quasiprobabilities from the perspective of the system}
\label{sec:SystemKDQ}

In \Sref{sec:QP}, we have introduced the stochastic instances $w$ and $q$, along with their respective quasiprobabilities, \eref{eq:QP_wc} and~\eref{eq:QP_qA}. These have been derived from the perspective of the environment's particles. Due to the energy-preserving condition $[U,H_S + H_A]$, we also have $u_S = -u_A$, such that it is equally possible to derive analogous expressions of coherent work and incoherent heat from the point of view of the quantum system.

In agreement with the first law of thermodynamics \eref{eq:first_law}, the stochastic instances $w_S$ and $q_S$ are both equal to $u_S$, as defined in~\eref{eq:def_uS}. 
Therefore, in analogy with~\eref{eq:QP_wc} and~\eref{eq:QP_qA}, the KDQs $\mathfrak{q}_n( w_{S}(\ell_{\rm in},\ell_{\rm fin}))$ and $\mathfrak{q}_n( q_S(\ell_{\rm in},\ell_{\rm fin}) )$ for the coherent work and incoherent heat assume the following expressions:
\begin{eqnarray*}
&&\mathfrak{q}_n\left( w_S \right) = 
\widetilde{\lambda}{\rm Tr}\left[ U^{\dagger} ( \Pi^{S}_{\ell_{\rm fin}} \otimes \mathbb{I}_A) U (\Pi^{S}_{\ell_{\rm in}} \otimes \mathbb{I}_A) \rho_S^{(n-1)} \chi_A \right],\\
&&\mathfrak{q}_n\left( q_S  \right) = 
{\rm Tr}\left[ U^{\dagger} (\Pi^{S}_{\ell_{\rm fin}} \otimes \mathbb{I}_A) U (\Pi^{S}_{\ell_{\rm in}} \otimes \mathbb{I}_A)  \rho_S^{(n-1)}  \rho_{A}^{\rm th} \right].
\end{eqnarray*}
It is worth noting that, as in the case of the KDQs defined from the perspective of the environment, the expressions for $w_S$ and $q_S$ only differ in the environment initial state, which is $\chi_A$ in the former case and $\rho_{A}^{\rm th}$ in the latter. Furthermore, the two KDQs are related to the quasiprobability of $u_S$, \eref{eq:QP_uS}, by
\begin{equation}
\mathfrak{q}_n( w_S) = \mathfrak{q}_n( u_S) - \mathfrak{q}_n( q_S).
\label{eq:KDQsystem}
\end{equation}
The KDQs $\mathfrak{q}_n( q_S)$ and $\mathfrak{q}_n( w_S)$ are, however, slightly different in nature: in the case of $q_S$, from ${\rm Tr}[\rho_{A}^{\rm th}] = 1$, it follows that $\sum_{(\ell_{\rm in},\ell_{\rm fin})} \mathfrak{q}_n( q_S(\ell_{\rm in},\ell_{\rm fin}) ) = 1$. As for $w_S$, instead, ${\rm Tr}[\chi_{A}] = 0$ implies $\sum_{(\ell_{\rm in},\ell_{\rm fin})} \mathfrak{q}_n( w_S(\ell_{\rm in},\ell_{\rm fin}) ) = 0$. While this may look odd, the quasiprobabilities $\mathfrak{q}_n( w_S)$ possess an operational and practical meaning in light of \eref{eq:KDQsystem}.

\section{Analytic expressions in the qubit-qubit case study}\label{sec:appC}

Let us consider the case study with two qubits described in \Sref{sec:Analyticresults}. In this Appendix we are going to give the analytical expression of the quasiprobabilities entering the distribution of internal energy variations for the quantum system and the environment particles, as well of the coherent work and incoherent heat. Notice that the expressions related to the internal energy variations are exact (i.e., not restricted to the small-$\tau$ limit), while the ones concerning the coherent work and incoherent heat are only meaningful for $\tau$ small enough that the unitary operator $U(\tau)$ is well approximated by its second-order expansion in $\tau$, and we can apply the second-order Baker-Campbell-Hausdorff series expansion.

\subsection{Quasiprobabilities}

We start from the real and imaginary parts of the quasiprobability $\mathfrak{q}_n\left( u_S(\ell_{\rm in},\ell_{\rm fin})\right)$~\eref{eq:QP_uS} with $\ell_{\rm in},\ell_{\rm fin}=0,1$, in the resonant case, for $\omega_S = \omega_A = \omega$. For the sake of a simplified notation, here we use the symbol $\mathfrak{q}_n( u_S^{\ell_{\rm fin}\ell_{\rm in}})$, where $u_S^{\ell_{\rm fin}\ell_{\rm in}} \equiv E^{S}_{\ell_{\rm fin}} - E^{S}_{\ell_{\rm in}}$ with $E^{S}_0 = \hbar \, \omega/2$ and $E^{S}_1 = -\hbar \, \omega/2$. From now on, throughout this Appendix we set also $\hbar = 1$. We find the following:
\begin{equation}\label{eq:real_parts_KDQ}
\cases{	
        \mathfrak{Re} \left[ \mathfrak{q}_n \left( u_S^{00}\right)\right] = 
        \frac{\rho_{11}^{(n)}}{Z} \big(  e^{\frac{\beta \, \omega}{2}} \cos^{2}(\varphi)+e^{-\frac{\beta \, \omega}{2}} \big) - \frac{J_2^{(n)}}{2}\sin(2\varphi) 
        \\
        \mathfrak{Re}\left[\mathfrak{q}_n\left( u_S^{01}\right)\right] = 
        \frac{1-\rho_{11}^{(n)}}{Z} e^{-\frac{\beta \, \omega}{2}} \sin^{2}(\varphi)-\frac{J_2^{(n)}}{2}\sin(2\varphi) 
        \\
        \mathfrak{Re}\left[\mathfrak{q}_n\left( u_S^{10}\right)\right] = \frac{\rho_{11}^{(n)}}{Z} e^{\frac{\beta\, \omega}{2}} \sin^{2}(\varphi) + \frac{J_2^{(n)}}{2}\sin(2\varphi) 
        \\
        \mathfrak{Re}\left[\mathfrak{q}_n\left( u_S^{11}\right)\right] = \frac{(1-\rho_{11}^{(n)})}{Z} \big( e^{-\frac{\beta \, \omega}{2}}\cos^{2}(\varphi) + e^{\frac{\beta \, \omega}{2}} \big) + \frac{J_2^{(n)}}{2}\sin(2\varphi)
}
\end{equation}
and
\begin{equation}\label{eq:imaginary_parts_KDQ}
\cases{
        \mathfrak{Im}\left[\mathfrak{q}_n\left( u_S^{00}\right)\right] =
        \frac{J_1^{(n)}}{2}\sin(2\varphi)
        \\
        \mathfrak{Im}\left[\mathfrak{q}_n\left( u_S^{01}\right)\right] = -\frac{J_1^{(n)}}{2}\sin(2\varphi) 
        \\
        \mathfrak{Im}\left[\mathfrak{q}_n\left( u_S^{10}\right)\right] = -\frac{J_1^{(n)}}{2}\sin(2\varphi) 
        \\
        \mathfrak{Im}\left[\mathfrak{q}_n\left( u_S^{11}\right)\right] = 
        \frac{J_1^{(n)}}{2}\sin(2\varphi)
}
\end{equation}
with $u_S^{00} = 0$, $u_S^{01} = \omega$, $u_S^{10} = -\omega$, $u_S^{11} = 0$. In~\eref{eq:real_parts_KDQ}-\eref{eq:imaginary_parts_KDQ}, $\varphi \equiv g \, \tau$, $J_1^{(n)} \equiv \lambda \,\mathfrak{Re}[\rho_{12}^{(n)}]$ and $J_2^{(n)} \equiv 
\lambda \,\mathfrak{Im}[\rho_{12}^{(n)}]$, and as usual $Z = e^{\, \beta \, \omega /2} + e^{-\beta \, \omega /2}$. Notice that quantum coherences of both the quantum system and the environment particle affect together the quasiprobabilities via the terms $J$'s, which are identically zero if either the system or the environment particle has no coherence. All the coherence-dependent terms develop in time proportionally to $\sin (g \, \tau)$, implying that, for short times, the effects they bring increase with time.

Since in this section we are considering the resonant case, we determine also the KDQ of both the incoherent heat and coherent work. However, it is worth recalling that the latter are valid only under the assumption that the collision time $\tau$ is small, as mentioned at the beginning of this section. Regarding the KDQ of the incoherent heat, they read as 
\begin{equation}
\label{eq:qqs}
\cases{
        \mathfrak{q}_n\left( q_S^{00}\right) = 
        \frac{\rho_{11}^{(n)}}{Z} \Big(  e^{\frac{\beta \, \omega}{2}} \cos^{2}(\varphi)+e^{-\frac{\beta \, \omega}{2}} \Big)  
        \\
        \mathfrak{q}_n\left( q_S^{01}\right) = 
        \frac{1-\rho_{11}^{(n)}}{Z} e^{-\frac{\beta \, \omega}{2}} \sin^{2}(\varphi) 
        \\
        \mathfrak{q}_n\left( q_S^{10}\right) = \frac{\rho_{11}^{(n)}}{Z} e^{\frac{\beta \, \omega}{2}} \sin^{2}(\varphi)  
        \\       
        \mathfrak{q}_n\left( q_S^{11}\right) = \frac{(1-\rho_{11}^{(n)})}{Z} \Big( e^{-\frac{\beta \, \omega}{2}}\cos^{2}(\varphi) + e^{\frac{\beta \, \omega}{2}} \Big) 
}
\end{equation}
where $q_S^{00} = 0$, $q_S^{01} = \omega$, $q_S^{10} = -\omega$, $q_S^{11} = 0$. These quasiprobabilities are purely real as expected, since they refer to the process of a simple interaction with a thermal bath. On the other hand, the real and imaginary parts of coherent work exchanges (again from the perspective of the system) are equal respectively to
\begin{equation}
\label{eq:qus}
\cases{
        \mathfrak{Re}\left[\mathfrak{q}_n\left( w_S^{00}\right)\right] = 
        - \frac{J_2^{(n)}}{2}\sin(2\varphi) 
        \\
        \mathfrak{Re}\left[\mathfrak{q}_n\left( w_S^{01}\right)\right] = 
        -\frac{J_2^{(n)}}{2}\sin(2\varphi) 
        \\
        \mathfrak{Re}\left[\mathfrak{q}_n\left( w_S^{10}\right)\right] =  \frac{J_2^{(n)}}{2}\sin(2\varphi) 
        \\
        \mathfrak{Re}\left[\mathfrak{q}_n\left( w_S^{11}\right)\right] =  \frac{J_2^{(n)}}{2}\sin(2\varphi)
}
\quad
\cases{
        \mathfrak{Im}\left[\mathfrak{q}_n\left( w_S^{00}\right)\right] = \frac{J_1^{(n)}}{2}\sin(2\varphi)
        \\
        \mathfrak{Im}\left[\mathfrak{q}_n\left( w_S^{01}\right)\right] = -\frac{J_1^{(n)}}{2}\sin(2\varphi) 
        \\
        \mathfrak{Im}\left[\mathfrak{q}_n\left( w_S^{10}\right)\right] = -\frac{J_1^{(n)}}{2}\sin(2\varphi) 
        \\
        \mathfrak{Im}\left[\mathfrak{q}_n\left( w_S^{11}\right)\right] = 
        \frac{J_1^{(n)}}{2}\sin(2\varphi)
}
\end{equation}
where $w_S^{00} = 0$, $w_S^{01} = \omega$, $w_S^{10} = -\omega$, $w_S^{11} = 0$. Not surprisingly, the quasiprobabilities of $q_S$ and or $w_S$ add up element by element to give those of $u_S$, which is another expressions of the first law of thermodynamics.

\subsection{Internal energy variation and variance at resonance}
\label{sec:AppAveVar}

Here we provide the analytic expressions for the average of the system's internal energy variation $\delta E_S^{(n)} = \langle u_S^{(n)} \rangle$, and its variance $(\Delta u_S^{(n)})^2$, in the resonant case $\Delta=0$. As explained in \Sref{sec:1stlaw}, for $\Delta=0$, $\delta E_A = - \delta E_S$; moreover, $\Delta u_A^2 = \Delta u_S^2$. We get:
\begin{eqnarray}
\label{eq:SystemEnergyRes}
   \delta E_S^{(n)} &=& - \omega  \left( \rho_{11}^{(n)} - \frac{e^{-\beta \, \omega /2}}{Z} \right) \sin^2 (\varphi) - \omega \,  J_2^{(n)} \sin (2 \varphi) \,  
   \\
   (\Delta u_S^{(n)})^2 &=& 
    \omega^2  \frac{ \left( e^{-\beta \, \omega/2} + \sinh (\beta \, \omega /2) \, \rho_{11}^{(n)} \right) }{Z} \sin^2 (\varphi) +
    \nonumber
    \\
    &-& i \, \omega^2 \, J_1^{(n)} \sin (2 \varphi)
    - \Big(\delta E_S^{(n)} \Big)^2 .
\end{eqnarray}

\section*{References}
\bibliography{paperIOP}

\end{document}